\documentclass[manuscript,authorversion]{acmart}

\makeatletter
\renewcommand\@formatdoi[1]{\ignorespaces}
\makeatother
\usepackage{amsmath}
\usepackage{booktabs} 
\usepackage{verbatim}
\usepackage{graphics} 
\usepackage{epsfig} 
\usepackage{epstopdf}
\usepackage{subfigure}
\usepackage{balance}
\usepackage{comment}
\usepackage{amsfonts}
\usepackage{bm}
\usepackage{colortbl}
\usepackage{tabularx}
\usepackage{color}

\usepackage{xspace}
\usepackage{graphicx}    
\usepackage{url}         
\usepackage{algorithm}
\usepackage{algpseudocode}
\usepackage{placeins}
\usepackage{multirow}
\usepackage{enumitem}
\usepackage{leading}
\usepackage{pgfplots}
\usepackage{tikz}
\usepackage{xcolor}
\usepackage[figuresright]{rotating}
\usepackage{soul}
\usepackage{pifont}

\AtBeginDocument{%
  \providecommand\BibTeX{{%
    \normalfont B\kern-0.5em{\scshape i\kern-0.25em b}\kern-0.8em\TeX}}}

\setcopyright{acmcopyright}
\copyrightyear{2018}
\acmYear{2018}
\acmDOI{10.1145/1122445.1122456}

\acmConference[CSUR]{ACM Computing Survey}
\acmBooktitle{ACM Computing Survey}
\acmPrice{15.00}
\acmISBN{978-1-4503-XXXX-X/18/06}

\begin{document}

\title{Key Generation for Internet of Things: A Contemporary Survey}
\author{Weitao Xu}
\email{weitaoxu@cityu.edu.hk}
\affiliation{%
  \institution{City University of Hong Kong}
  \country{Hong Kong SAR China}
}

\author{Junqing Zhang}
\affiliation{%
  \institution{University of Liverpool}
  \country{UK}}
\email{junqing.zhang@liverpool.ac.uk}

\author{Shunqi Huang}
\affiliation{%
  \institution{Shenzhen University}
  \country{China}}
\email{huangshunqi2019@email.szu.edu.cn}

\author{Chengwen Luo}
\affiliation{%
  \institution{Shenzhen University}
  \country{China}}
\email{chengwen@szu.edu.cn}

\author{Wei Li}
\affiliation{%
  \institution{The University of Sydney}
  \country{Australia}
}
\email{weiwilson.li@sydney.edu.au}


\renewcommand{\shortauthors}{Xu et al.}

\begin{abstract}
Key generation is a promising technique to bootstrap secure communications for the Internet of Things (IoT) devices that have no prior knowledge between each other. In the past few years, a variety of key generation protocols and systems have been proposed. In this survey, we review and categorise recent key generation systems based on a novel taxonomy. Then, we provide both quantitative and qualitative comparisons of existing approaches. We also discuss the security vulnerabilities of key generation schemes and possible countermeasures. Finally, we discuss the current challenges and point out several potential research directions.
\end{abstract}



\maketitle

\section{Introduction}

\begin{figure}
	\begin{minipage}{0.48\linewidth}
		\centering
		\includegraphics[width=3.5in]{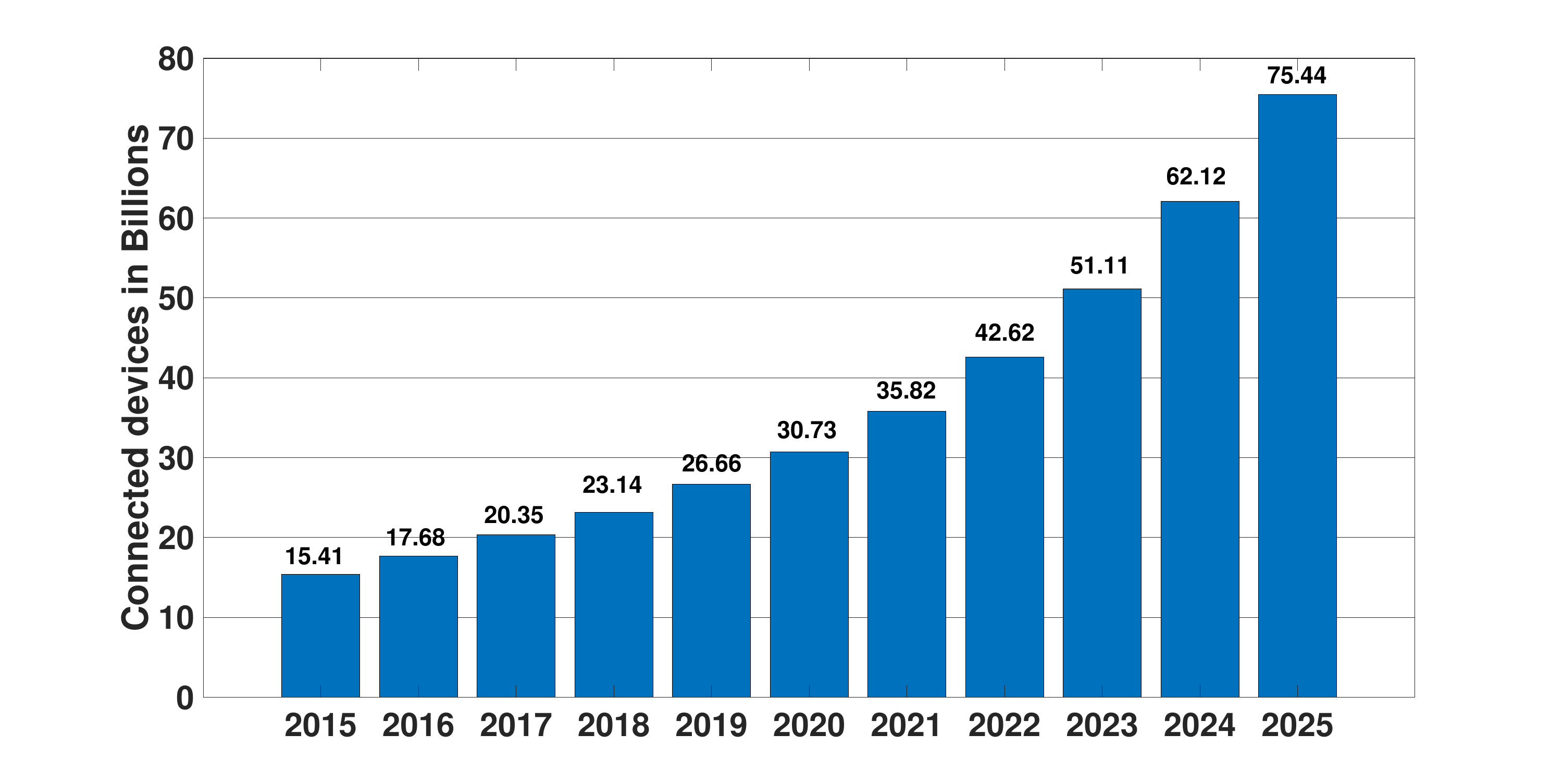}
    \caption{Number of IoT devices from 2015 to 2025~\cite{statista2018internet}.}
    \label{fig:iottrend}
	\end{minipage}
	\hfill
	\begin{minipage}{0.47\linewidth}
		\centering
		\includegraphics[width=3.2in]{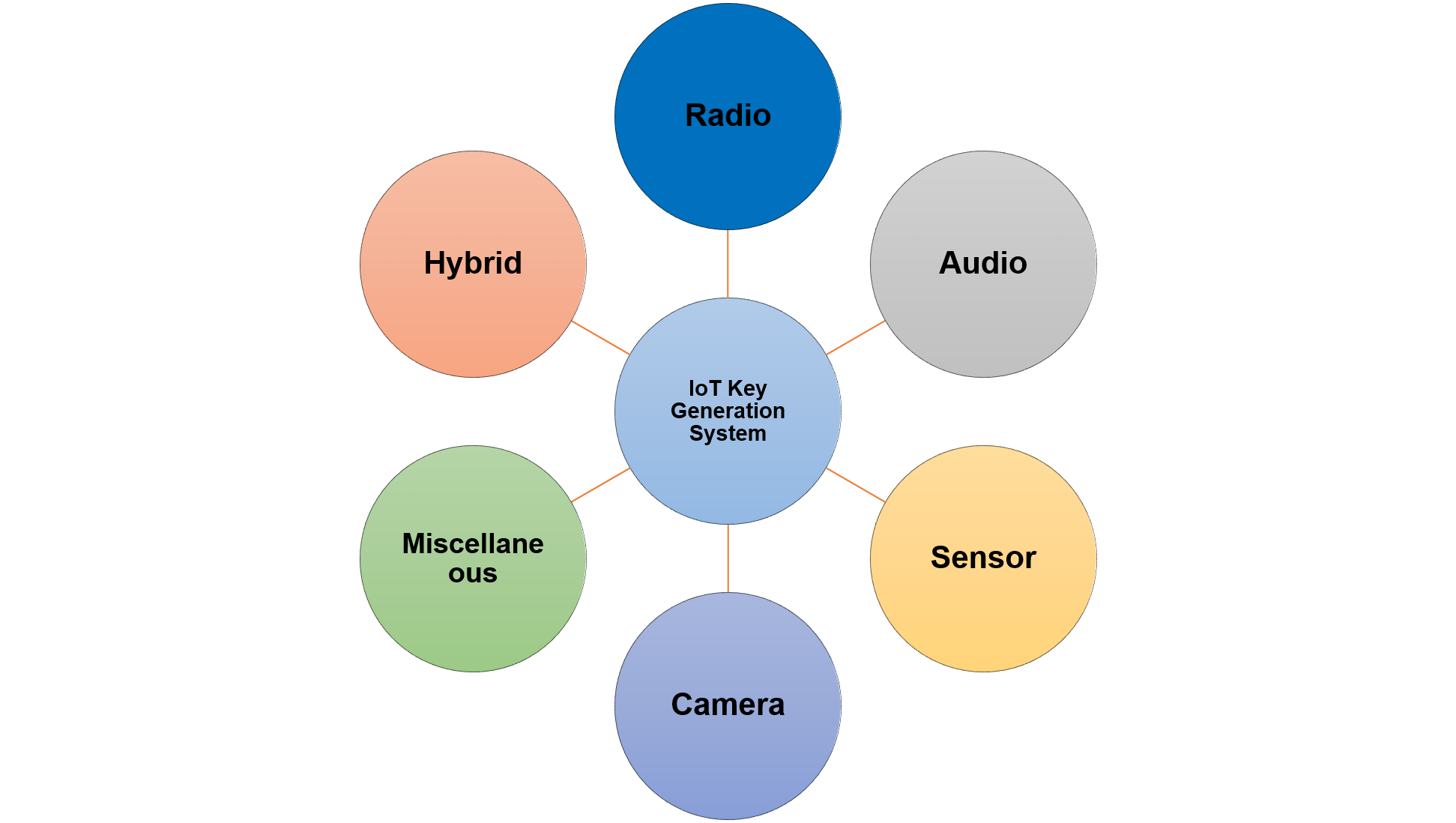}
		\caption{Taxonomy of our survey.}
		\label{fig:taxonomy}
	\end{minipage}
	\vspace{-0.3in}
\end{figure}

The last decades have witnessed the rapid growth of the Internet of Things (IoT) from a theoretical concept to a reality. A wide range of smart IoT devices have penetrated into our daily life. These devices can be broadly classified into three classes: portable/wearable devices such as smart watch and Fitbit, smart home/building devices such as Amazon Alexa and Google Assistant, general network devices such as Wi-Fi router and 5G end device. We are entering a new era where every thing/device will be connected together to form a smart world. Fig.~\ref{fig:iottrend} shows the growth of the number of IoT devices in the past decade as well as the predictions of the growth until 2025. It is estimated that the number of IoT devices connected to the Internet will surge to 75 billion by 2025~\cite{statista2018internet}.

With the prevalence of IoT devices, secure device-to-device (D2D) communication is becoming more and more crucial because IoT devices often need to be paired together for the purpose of file transferring, synchronisation, and data sharing. Cryptographic key agreement is a fundamental requirement for secure D2D communications to achieve confidentiality~\cite{xi2014keep}. Key generation, also called key agreement or establishment, refers to the process of generating the same cryptographic key between two devices that have no prior secret. Alternative terminologies, such as device association, device pairing, or device binding, have the same meaning but are adopted by other literature~\cite{chong2012usability}. Essentially, the underlying concept remains the same, i.e., to establish a secure communication channel among multiple IoT devices. In the following, we will use key generation, key agreement, key establishment and device pairing interchangeably in spite of the minor differences in some contexts~\cite{fomichev2017survey}. Broadly speaking, key generation can be divided into two classes: authenticated key generation and unauthenticated key generation. In the authenticated key generation, one of the communication parties can verify the identity of the other device. While in the unauthenticated key generation, they simply generate a pair of key without authenticating each other.  

Secure key establishment between two parties can be completed by public key cryptography (PKC)~\cite{shim2016survey}. PKC schemes require a public key infrastructure (PKI) and are computationally expensive as they usually rely on complicated mathematical problems, e.g., discrete logarithm algorithm.
Hence, PKC solutions may not be suitable to resource-constrained devices operating in pervasive environments due to both the absence of PKI and the required high computational overhead.
Another solution is pre-shared key (PSK) scheme. However, PSK scheme lacks scalability which makes them inappropriate especially in the cases of large-scale sensor deployments and mobile scenarios. 

The resource constraints and the absence of common trust infrastructure motivate researchers to seek alternative key generation approaches by exploring the features and functions integrated in IoT devices themselves. Existing key generation systems are based on a common principle: if multiple devices can observe a common signal (e.g, sound, temperature, or motion) from a specific channel, their observations can be used as materials to generate random keys. Prior researches have yielded a massive number of technically sound systems relying on various auxiliary out-of-band (OOB) channels such as visual channel~\cite{mccune2005seeing}, acoustic channel~\cite{schurmann2011secure} and sensing channel~\cite{mayrhofer2009shake}. 

The concept of IoT device pairing is not new and there already exist several literature surveying this area~\cite{Shahab2014survey,shehadeh2015survey,zhang2016review,fomichev2017survey}. However, with the advent of new wireless technologies such as Long Range communication technology (LoRa) and new hardware such as bio-sensors, many novel secure device pairing schemes have been proposed recently~\cite{xu2019lora,ruotsalainen2019experimental,ruiz2020IDIoT,roeschlin2018device}. Unfortunately, these surveys do not capture the recent advances in IoT device pairing. Therefore, the aim of our survey is to bridge this gap in the current literature. Moreover, the taxonomies used in previous surveys have some weaknesses. For example, Shahab et al.~\cite{Shahab2014survey} categorised device pairing systems into three classes: weak, public, and private channels. However, the difference between these three categories are unclear because bias is unavoidable when using subjective metrics. In a more recent survey~\cite{fomichev2017survey}, Fomichev et al. analysed secure pairing schemes from three aspects: physical channel, human-computer interaction (HCI) and application classes. However, when we review the literature we find this systematisation is not easy to use in practice because a large portion of systems actually lie in the intersection of different categories. For instance, many schemes belonging to application classes in~\cite{fomichev2017survey} also involve physical channel and HCI.

In comparison with prior works, our survey presents two novel contributions. First, our survey complements the previous surveys in terms of recent key generation approaches and systems. Second, our survey summarises and compares existing works in a new perspective. Specifically, we use a novel taxonomy to organise this survey, as shown in Fig.~\ref{fig:taxonomy}. The rationale for introducing this new taxonomy is to classify approaches by the hardware used by different approaches, contrasting previous surveys that were organised from the perspective of HCI~\cite{chong2014survey,fomichev2017survey}. We argue that a classification based on the hardware interface is an important complement to previous survey works, as the fundamental difference of materials or information used to generate keys is they originate from different hardware interfaces such as radio, audio, and sensor. Table~\ref{tab:surveycomparison} compares the coverage of our survey against previously published ones.  For topics well studied in the prior works, we only provide a brief summary in this paper for the benefit of the readers and for the sake of completeness, while the details are referred to those references for additional information. 
\begin{table*}[t]
\centering
\small
\caption{Comparison of prior surveys/reviews with this survey (~\ding{109}--none,~\ding{119}--moderate,~\ding{108}--comprehensive).}
\label{tab:surveycomparison}
\centering
\begin{tabular}{cccccccccc}
\toprule
                               & Year & Radio & Audio & Sensor & Camera & Miscellaneous & Hybrid & \begin{tabular}[c]{@{}c@{}}Security\\ Analysis\end{tabular} & \begin{tabular}[c]{@{}c@{}}Performance\\ Comparison\end{tabular} \\ \hline
\cite{kumar2009comparative}                             & 2009 & \ding{119}     & \ding{119}     & \ding{109}      & \ding{119}      & \ding{109}             & \ding{109}      & \ding{109}                                                           & \ding{108}                                                                \\ \hline
\cite{Shahab2014survey} & 2014 & \ding{108}     & \ding{108}     & \ding{119}      & \ding{108}      & \ding{109}             & \ding{109}      & \ding{108}                                                           & \ding{109}                                                                \\ \hline
\cite{chong2014survey}   & 2014 & \ding{108}     & \ding{108}     & \ding{108}      & \ding{108}      & \ding{109}             & \ding{109}      & \ding{109}                                                           & \ding{108}                                                                \\ \hline
\cite{shehadeh2015survey}                              & 2015 & \ding{119}     & \ding{109}     & \ding{109}      & \ding{109}      & \ding{109}             & \ding{109}      & \ding{109}                                                           & \ding{109}                                                                \\ \hline
\cite{zhang2016review}                        & 2016 & \ding{108}     & \ding{109}     & \ding{109}      & \ding{109}      & \ding{109}             & \ding{109}      & \ding{109}                                                           & \ding{109}                                                                \\ \hline
\cite{fomichev2017survey}                              & 2017 & \ding{108}     & \ding{108}     & \ding{108}      & \ding{108}      & \ding{119}             & \ding{109}      & \ding{109}                                                           & \ding{109}                                                                \\ \hline
Ours                           & 2020 & \ding{108}     & \ding{108}     & \ding{108}      & \ding{108}      & \ding{108}             & \ding{108}      & \ding{108}                                                           & \ding{108}                                                                \\ \bottomrule
\end{tabular}
\vspace{-0.2in}
\end{table*}

The rest of the paper is organised as follows. In Section~\ref{sec:IoTdevices}, we classify the state-of-the-art IoT devices by examining their features and functions and derive a taxonomy based on the analysis. Section~\ref{sec:keygeneration}  surveys representative key generation systems based on the used hardware and technologies. 
Then, a comparison of existing key generation schemes is given in Section~\ref{sec:performancecomparison}.
Section~\ref{sec:attack_countermeaures} reviews the security vulnerabilities of key generation schemes and discusses countermeasures. 
Current challenges and future directions are discussed in Section~\ref{sec:challenges_directions}. Finally, Section~\ref{sec:conclusion} concludes the paper.

\section{A Brief Survey of IoT devices}
\label{sec:IoTdevices}
In this section, we present a brief survey
of existing IoT devices/products. The aim of this section is not to provide a comprehensive analysis and categorisation of IoT devices, but to give readers a general idea regarding the features and resources that can be provided by different types of devices.  For a comprehensive survey of IoT systems, the readers are referred to~\cite{bansal2020iot,seneviratne2017survey}.

IoT devices are essentially smart devices that support internet connectivity and are able to communicate over the internet with other devices and provide a user with remote access to control the device according to their needs. These devices generally integrate sensing, processing, and communication to facilitate autonomous awareness of the context of a device. Wireless connectivity is introduced to allow sharing such context among IoT networks. Based on application scenarios, we classify the commonly used IoT devices into three categories: mobile/wearable devices, smart home/building devices, and network devices. Table~\ref{tab:iotdevices} presents the details of devices discussed in this paper.
\begin{table}[!t]
\centering
\small
\caption{A summary of common IoT devices.}
\label{tab:iotdevices}
\scalebox{0.70}{
\begin{tabular}{lllll}
\toprule
Category & Subcategory    & Example products     & Common Features   & Application Scenarios                                                                                                                                                                                                                                                      \\ \hline
\multirow{11}{*}{\begin{tabular}[c]{@{}l@{}}Portable/Wearable\\ Devices\end{tabular}}  & Smart phone    & \begin{tabular}[c]{@{}l@{}}Apple phone\\ Samsung phone\\ Huawei phone\end{tabular}             & \begin{tabular}[c]{@{}l@{}}Sensors: camera, accelerometer, gyroscope, geomagnetic sensor, barometer, \\ proximity sensor, ambient light sensor\\ Wireless Connectivity: Bluetooth 5.0, Wi-Fi, GPS, LTE, GSM, NFC, iBeacon\\ Microphone, Speaker: Yes\end{tabular}  &  \multirow{11}{*}{\begin{tabular}[c]{@{}l@{}}Health monitoring\\ Activity detection\\ Localisation\\ Entertainment \\ AR/VR\end{tabular}}   \\ \cline{2-4} 
                                                                                       & Smart watch    & \begin{tabular}[c]{@{}l@{}}Apple Watch\\ Huawei Watch\end{tabular}                        & \begin{tabular}[c]{@{}l@{}}Sensors: accelerometer, gyroscope, geomagnetic sensor, optical heart rate sensor,\\ ambient light sensor, air pressure sensor, capacitive sensor, \\ Wireless Connectivity: Bluetooth, BLE, NFC, GPS\\ Microphone, Speaker: Yes\end{tabular} \\ \cline{2-4} 
                                                                                       & Smart glass    & \begin{tabular}[c]{@{}l@{}}Google Glass\\ Vuzix Smart Glass\\ Magic Leep One\end{tabular} & \begin{tabular}[c]{@{}l@{}}Sensors: camera,  accelerometer, gyroscope, geomagnetic sensor, \\ multi-touch gesture touchpad, IR eye tracking, depth sensor\\ Wireless Connectivity: Bluetooth, Wi-Fi\\ Microphone, Speaker: Yes\end{tabular}                             \\ \cline{2-4} 
                                                                                       & Smart clothing & \begin{tabular}[c]{@{}l@{}}Mercury Intelligent Jacket\\ Nadi X Yoga Pants\end{tabular}    & \begin{tabular}[c]{@{}l@{}}Sensors: accelerometer, internal and external temperature sensor\\ Wireless Connectivity: Bluetooth\\ Output component: a smart thermostat, haptic feedback (vibration)\\ Microphone, Speaker: No\end{tabular}                               \\ \cline{2-4} 
                                                                                       & Smart shoes    & Nike Adapt BB 2.0 Shoes                                                                   & \begin{tabular}[c]{@{}l@{}}Sensors: accelerometer, gyroscope, capacitive touch controller\\ Wireless Connectivity: BLE\\ Microphone, Speaker: No\end{tabular}                                                                                                           \\ \hline
\multirow{8}{*}{\begin{tabular}[c]{@{}l@{}}Smart home/Building\\ Devices\end{tabular}} & Smart speaker  & \begin{tabular}[c]{@{}l@{}}Google Assistant\\ Amazon Echo\end{tabular}                    & \begin{tabular}[c]{@{}l@{}}Sensors: No\\ Wireless Connectivity: Bluetooth 5.0, Wi-Fi\\ Microphone, Speaker: Yes\end{tabular}  & \multirow{8}{*}{\begin{tabular}[c]{@{}l@{}} Surveillance\\ Intrusion detection\\ Environmental monitoring\\ Smart meter \\ Building automation \\ Smart parking\end{tabular}}                                                                                                                                          \\ \cline{2-4} 
                                                                                       & Smart display  & \begin{tabular}[c]{@{}l@{}}Google Nest Hub\\ Lenovo Smart Display\end{tabular}            & \begin{tabular}[c]{@{}l@{}}Sensors: capacitive touch screen, ambient light sensor\\ Wireless Connectivity: Bluetooth 5.0, Wi-Fi\\ Microphone, Speaker: Yes\end{tabular}                                                                                                 \\ \cline{2-4} 
                                                                                       & Smart bulb     & \begin{tabular}[c]{@{}l@{}}Philips Hue\\ Lifx Z LED Strip\end{tabular}                    & \begin{tabular}[c]{@{}l@{}}Sensors: No\\ Wireless Connectivity: Bluetooth, Wi-Fi, ZigBee\\ Microphone, Speaker: No\end{tabular}                                                                                                                                         \\ \cline{2-4} 
                                                                                       & Miscellaneous  & \begin{tabular}[c]{@{}l@{}}Bosch BCC50 \\ Google Nest Secure System\end{tabular}          & \begin{tabular}[c]{@{}l@{}}Sensors: thermostat, motion sensor, vibration sensor, humidity sensor,\\ PIR sensor, proximity sensor, etc\\ Wireless connectivity: Bluetooth, Wi-Fi, NFC, ZigBee, Cellular\\ Microphone, Speaker: depending on specific device\end{tabular} \\ \hline
\multirow{6}{*}{Network Devices}                                       & Wi-Fi          & \begin{tabular}[c]{@{}l@{}}Google Nest Wifi \\ TP-Link Deco X60\end{tabular}              & \multirow{6}{*}{\begin{tabular}[c]{@{}l@{}}Sensors: No\\ Wireless connectivity: Yes, depending on the specific communication technology\\ Microphone, Speaker: No\end{tabular}}                                                                                       & \multirow{9}{*}{\begin{tabular}[c]{@{}l@{}} Smart city\\ Smart farm \\ Smart transportation \end{tabular}}  \\ \cline{2-3}
                                                                                       & ZigBee         & \begin{tabular}[c]{@{}l@{}}Sangsung SmartThings Hub\\ Hive Hub\end{tabular}               &                                                                                                                                                                                                                                                                         \\ \cline{2-3}
                                                                                       & LoRa           & \begin{tabular}[c]{@{}l@{}}Arduino MKR WAN 1300\\ Cisco IR 910\end{tabular}               &                                                                                                                                                                                                                                                                         \\ \cline{2-3}
                                                                                       & Bluetooth      & Wink Hub                                                                                  &                                                                                                                                                                                                                                                                         \\ \cline{2-3}
                                                                                       & RFID           & LANMU RFID Card Reader                                                                    &                                                                                                                                                                                                                                                                         \\ \cline{2-3}
                                                                                       & 5G             & HTC 5G Hub                                                                                &                                                                                                                                                                                                                                                                         \\ \bottomrule
\end{tabular}
}
\vspace{-0.2in}
\end{table}

\textbf{Portable/wearable devices.} Portable/wearable devices are smart electronic devices that are worn close to and/or on the surface of the skin or carried by a user. These devices are often called wearables for short and examples of such devices include smart glass, smart watch, and smart clothing. Here we also classify mobile phone into this category because many so-called mobile devices are not mobile themselves, it is the host that carries these devices is mobile. As shown in Table~\ref{tab:iotdevices}, these devices are usually equipped with a variety of sensors that can monitor, detect, analyse and transmit information about user's context such as location, motion, heart rate etc. They also have built-in microphone, speaker and different wireless connectivity functionalities. Therefore,  compared to the other two types of devices, wearable devices can provide more information to generate keys. Indeed, a large portion of existing key generation systems are based on wearable devices~\cite{shangaudiokey,xie2018genewave,shen2018shake,xu2016walkie,xu2017gait,jin2015magpairing}.

\textbf{Smart home/building devices.} There are a variety of smart home/building devices ranging from smart sensors (e.g., proximity sensor and vibration sensor) to smart appliances (e.g., Bosch BCC50). Although several smart home applications have built-in sensors, a large majority of them do not have sensors, microphone and speaker. Key generation systems for smart home/building devices usually exploit the common context information that can be measured by heterogeneous sensors~\cite{miettinen2014context,han2018you}. 

\textbf{Network devices.} Network devices are physical devices that are needed on a computer network to communicate and interact with hardware. Network device usually do not have built-in sensors, microphone and speaker. Moreover, each network device is usually equipped with single wireless communication interface. Key generation systems for these devices are usually based on wireless channel physical layer characteristics such as Received Signal Strength Indicator (RSSI), Channel State Information (CSI) etc.

The variety of IoT devices compounds the complexity of key generation systems for two reasons. First, the hardware capabilities available, such as wireless radio interfaces, sensing functionality, and computing capacity, vary greatly across different platforms as can be seen in Table~\ref{tab:iotdevices}. Secondly, the patterns of interaction between the systems and between human operators and machines have become more complex. Therefore, to identify existing studies and encourage further work on this subject, we are motivated to propose a new taxonomy to systematise information in this area. As shown in Fig.~\ref{fig:taxonomy}, we categorise existing key generation systems based on the adopted hardware interface. The reason of introducing this new taxonomy is that the hardware interface manifests the fundamental difference between different material/information used to generate keys. With the proposed taxonomy, users can easily identify the most suitable method for their application scenarios. For example, for wearable devices which are equipped with rich sensors, users can select Inertial Measurement Unit (IMU) sensor-based or audio-based key generation techniques. For network devices which are equipped with wireless modules only, the radio-based key generation system is the best option.

\section{Key Generation Systems for IoT Devices}
\label{sec:keygeneration}
In this section, we survey the key generation systems for IoT devices based on the taxonomy in Fig.~\ref{fig:taxonomy}. Specifically, we discuss key generation systems based on the hardware or the channel used to collect information for generating keys. We divide them into six categories: radio, audio, IMU sensor, camera, miscellaneous hardware and hybrid approaches. We will survey the representative works of each category in turn. 

\subsection{Overview of Threats}
Before discussing key generation systems, we first give a brief overview of potential attacks. This is because security systems always come along with attacks. With attacks in mind, readers can better understand the vulnerabilities of different approaches. Without loss of generality, we adopt the notations commonly used in this field: Alice and Bob represent two legitimate devices that aim to generate the same key, and Eve represents an attacker trying to obtain the same key via different types of attacks.  The goal of Eve is always to undermine the device pairing system and she can launch attacks on authenticity, confidentiality and integrity. The formulation of a detailed adversary model is beyond the scope of this article. In this paper, we focus on some common attacks such as Man-in-the-Middle (MITM) attack, replay attack, eavesdropping attack. These attacks will have different degrees of threats to different schemes described below. A detailed discussion of attacks and countermeasures is given in Section~\ref{sec:attack_countermeaures}.

\subsection{Radio}\label{sec:radio}
Radio communication modules have been pre-installed in many consumer electronics. For example, all the smartphones have cellular, Wi-Fi and Bluetooth connectivities. Fitbits are usually equipped with Bluetooth. Most of the IoT wireless techniques operate at the unlicensed industrial, scientific, and medical (ISM) band. For instance, Wi-Fi runs at 2.4 GHz or 5 GHz; ZigBee and Bluetooth work at 2.4 GHz; LoRa operates at the sub GHz and its frequency plan depends on the regions, e.g., 868 MHz in EU and 902-928 MHz in the US.

\begin{figure*}[!t]
	\centering
	\subfigure[Channel reciprocity-based]{
		\includegraphics[height=1.33in]{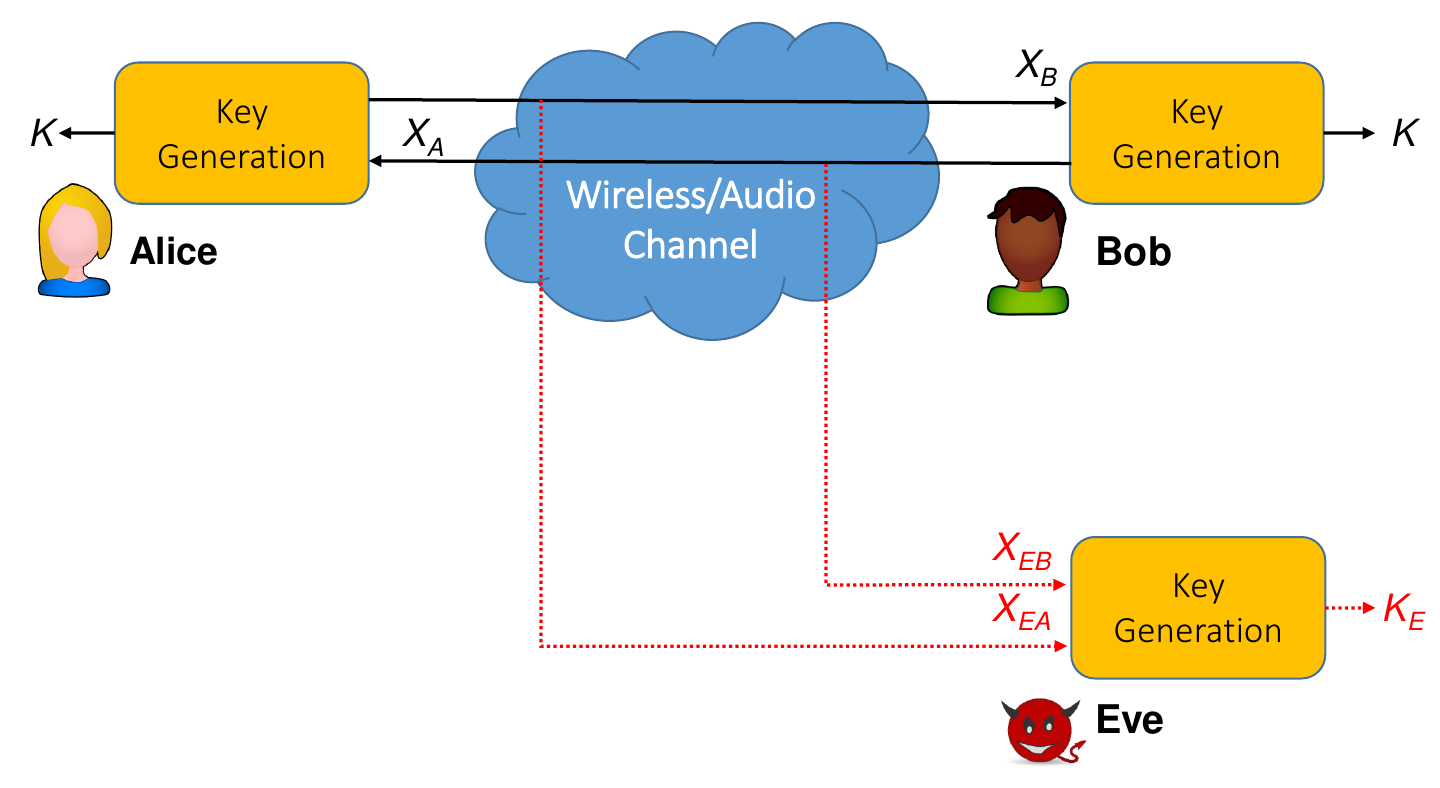}
		\label{fig:channelreciprocitybased}} \hspace{0.2cm}
		\subfigure[Proximity-based]{
		\includegraphics[height=1.2in]{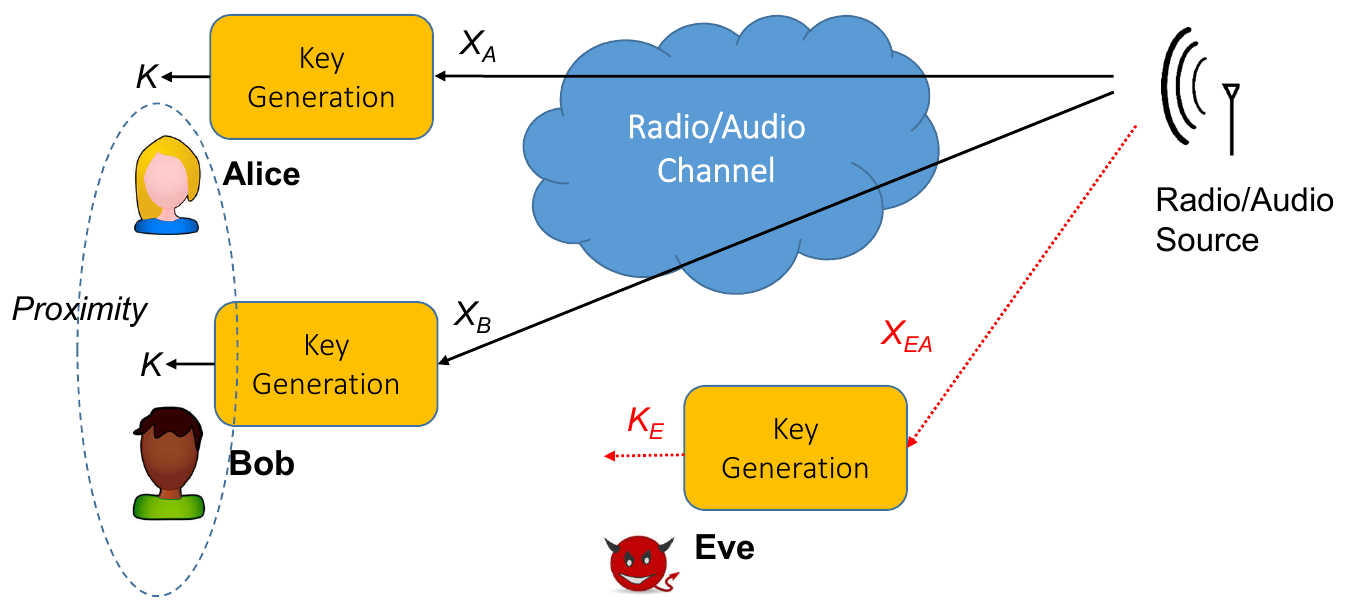}
		\label{fig:proximitybased}}
	\caption{System model for radio and audio-based key generation.}
	\label{fig:model}
	\vspace{-0.2in}
\end{figure*}

Radio communications are used to exchange information between devices. At the same time, they can also be exploited for generating secret keys.
Key generation with radio communications has received extensive interests in the last decades~\cite{zhang2016review,zhang2019physical}.
Depending on the locations of Alice and Bob, radio-based key generation can be categorised into channel reciprocity-based and proximity-based schemes, as illustrated in Fig.~\ref{fig:model}.
In the channel reciprocity-based scheme, Alice and Bob wish to generate the same key from their common wireless channel and prevent the key secure against eavesdroppers.
On the other hand, in the proximity-based scheme, Alice and Bob are located very close to each other and receive signals from a common third user.

\subsubsection{Principle}
Radio-based Key generation, also referred to as physical layer key generation, relies on three principles, namely temporal variation, channel reciprocity and spatial decorrelation, which will be explained in detail below.
\begin{itemize}
	\item \textit{Temporal Variation.}
The signal propagation is subject to reflection, refraction and scattering and hence the channel effects will be temporally varying in a mobile environment. These effects will be unpredictable and their randomness can be extracted as cryptographic keys.
	\item \textit{Channel Reciprocity.}
When the uplink and downlink transmissions operate at the same carrier frequency, the channel effects at both ends of the link will be reciprocal. This feature ensures that Alice and Bob can obtain highly correlated channel measurements and generate the same keys.
	\item \textit{Spatial Decorrelation.}
When Eve is located more than half-wavelength away from either of the legitimate users, she experiences uncorrelated channel effects according to the communication theory. This property is essential to guarantee that the keys generated by Alice and Bob cannot be guessed by Eve hence the keys are secure.
\end{itemize}
The radio-based key generation principles have been modelled and analysed in~\cite{zhang2017key} and experimentally validated in~\cite{zhang2016experimental}. There has been extensive work evaluating these principles, e.g., by using Wi-Fi~\cite{zhang2016experimental}, ultrawideband (UWB)~\cite{hamida2010empirical}.

\subsubsection{Protocol}
\begin{figure}[!t]
	\centering
		\includegraphics[width=2.8in]{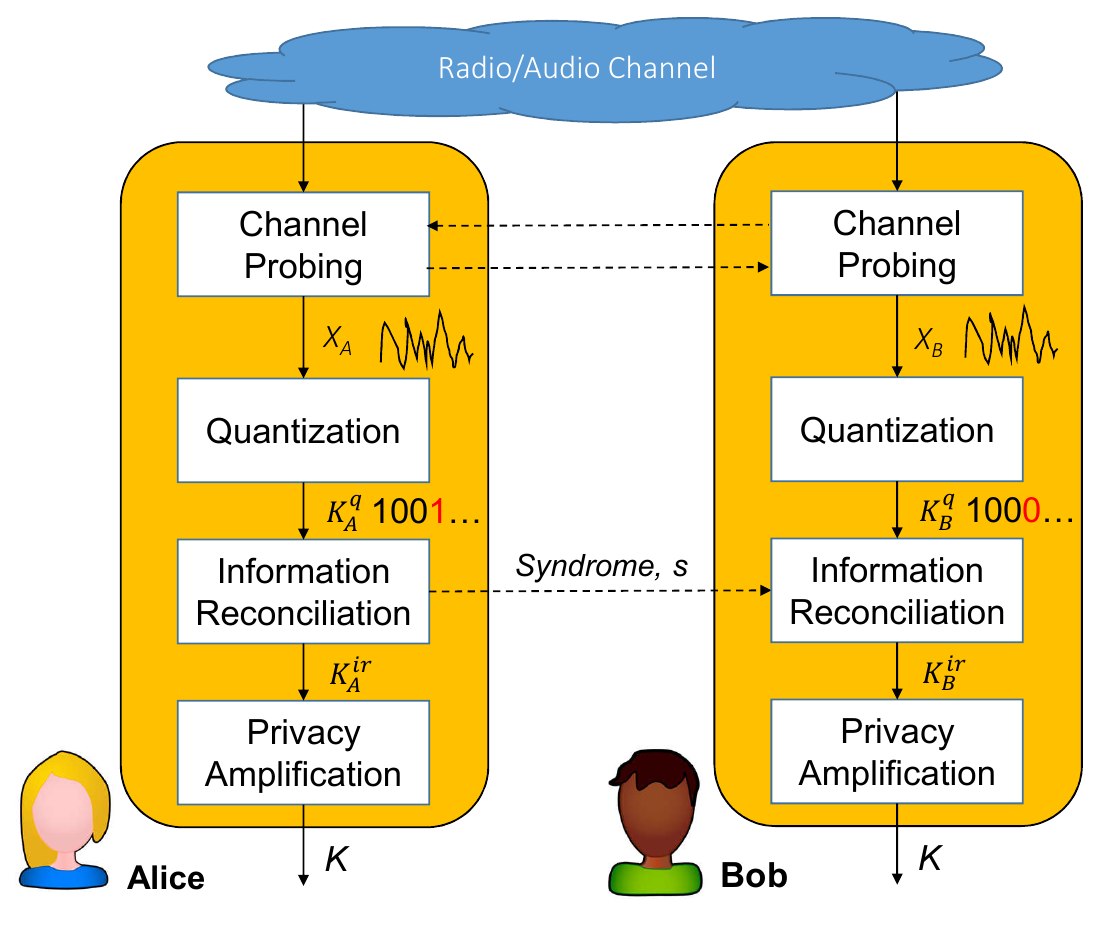}
	\caption{Work flow of channel reciprocity based key generation.}
	\label{fig:protocol}
	\vspace{-0.2in}
\end{figure}

As shown in Fig.~\ref{fig:protocol}, a typical channel reciprocity-based key generation protocol consists of the following four steps: channel probing, quantization, information reconciliation and privacy amplification.

\paragraph{Channel Probing}
This step requires bidirectional transmissions between Alice and Bob. In particular, Alice first transmits a signal to Bob who will measure the channel information via some channel parameters, e.g., RSSI, channel impulse response (CIR) and channel frequency responses (CFR). Bob will then reply a signal to Alice who will measure the same parameter. This completes a pair of measurements. Alice and Bob keep probing until they collect sufficient measurements, $X_A$ and $X_B$, respectively.

The selection of the channel parameter depends on the adopted wireless technologies. While RSSI is almost provided in all the wireless standards, CIR and CFR are only available in wideband systems.
For example, UWB systems are able to estimate the CIR~\cite{wilson2007channel,bulenok2016experimental} while IEEE 802.11 Orthogonal Frequency-Division Multiplexing (OFDM) can obtain the CFR~\cite{liu2013fast,xi2014keep,zhang2016efficient,zhang2019design}.

\paragraph{Quantization}
As cryptographic algorithms require binary input, the analog measurements, $X_u$, should be converted into binary sequence $K_u$, which can be completed by comparing $X_u$ to thresholds and assign bit sequences. 
The thresholds can be determined based on the mean value and variance as well as cumulative distribution functions (CDF).

\textbf{Mean and standard deviation-based quantizer} ~\cite{mathur2008radio,jana2009effectiveness} chooses the thresholds  based on the mean value and the variance of the channel measurements. Specifically, the bit assignment can be completed as follows:
\begin{align}
	K_u(i) = \begin{cases}
	1, \; \text{when}\; X_u(i) > \mu_u + \alpha\times \sigma \\
	0,  \; \text{when}\; X_u(i) < \mu_u - \alpha\times \sigma
	\end{cases}
\end{align}
where $\mu_u$ is the mean value and $\sigma$ is the standard deviation. This quantizer is very easy to implement. However, it is not robust to burst error.

\textbf{CDF-based quantizer}~\cite{patwari2010high} calculates the thresholds based on the CDF of the channel measurements, $F(x)$. CDF-based quantizer can be implemented as a multi-bit quantizer. The thresholds can be calculated as 
\begin{align}
	\eta_n = F^{-1}(\frac{n}{2^{QL}})
\end{align}
where $n = 1, 2, ..., 2^{QL} - 1$, $QL$ is the quantization level, and $F^{-1}(\cdot)$ is the inverse function of the CDF. We can assign Gray code $b_n$ to the range $[\eta_{n-1}, \eta_n]$. Finally, the keys can be generated as
\begin{align}
	K_u(i,QL) =  b_n, \text{when} \; \eta_{n-1} \leq X_u(i) < \eta_{n}.
\end{align}
Compared to the mean and standard deviation-based quantizer, the CDF quantizer can generate multiple bits from one measurement, which is more efficient. However, it is more complex as it requires to calculate the CDF and its inverse function.
A detailed comparison between different quantizers can be found in~\cite{guillaume2014fair,zenger2015security}.

\paragraph{Information Reconciliation}
There will probably be key mismatch between Alice and Bob after the quantization, due to the channel variation, hardware asymmetry and noise. The mismatch can be quantified by the key disagreement rate (KDR), given as
\begin{align}
	KDR = \frac{\sum_{i}|K_A(i) - K_B(i)|}{l_K}
\end{align}
where $l_K$ is the key length.

Information reconciliation is thus employed to make Alice and Bob agree on the same keys, which can be achieved using error correction codes (ECCs). Secure sketch~\cite{dodis2008fuzzy} is a popular algorithm, which has been widely used in the wireless key generation to correct the mismatch~\cite{wang2011fast,zhang2016efficient,zhang2018channel}. Alice first randomly selects a codeword $c$ from the codeset of ECC. She calculates the syndrome by $s = c \oplus  K_A$ and sends it to Bob.
Assuming Bob receives the syndrome successfully without error, he then obtains $c_B = s \oplus  K_B$. If the hamming distance between $c_B$ and $c$ is within the correction capacity of the ECC, Bob will be able to decode $c$ from $c_B$. Finally, Bob can calculate $K_B^{ir} = c \oplus  s$, which should be the same as the key generated at Alice. 

\paragraph{Privacy Amplification}
There will be information leaked to eavesdroppers during the information reconciliation, hence privacy amplification is adopted to mitigate the threat of information leakage. In the literature, this is done by extractor~\cite{wang2011fast}, universal hashing functions~\cite{jana2009effectiveness}, or cryptographic hash functions~\cite{zhang2010mobility}. After this step, key generation is completed. Alice and Bob get the same secure key and they can use this key with symmetric encryption algorithms such as AES-128 to secure their subsequent communications.

\subsubsection{Channel Reciprocity-based Approaches}\label{sec:reciprocity_radio}
Channel reciprocity-based key generation has received extensive research interests and been applied with several wireless technologies, such as ZigBee, Wi-Fi, LoRa, etc.
As shown in Fig.~\ref{fig:model}(a), Alice and Bob are located with a certain distance between them. They will then leverage the reciprocal channel between them for key generation. Below, we introduce several representative systems for each wireless technology.

\paragraph{ZigBee}
ZigBee is a popular communication technology for wireless sensor networks and wireless body area networks. It uses the IEEE 802.15.4 as the physical and Media Access Control (MAC) layer protocols. ZigBee operates at the 2.4~GHz and has a communication range about 100 meters.
RSSI is available in ZigBee systems.

To the best of the authors' knowledge, Aono~et al.~\cite{aono2005wireless} proposed the first ZigBee-based key generation protocol in 2005, which is also the first practical key generation system. The authors designed an electronically steerable parasitic array radiator (ESPAR) antenna, which can be dynamically configured to introduce channel fluctuation. Patwari~et al.~\cite{patwari2010high} proposed high-rate uncorrelated bit extraction (HRUBE), which is a framework incorporating interpolating, transforming for decorrelation and multi-bit quantization.
They constructed a testbed using the TI CC2420 sensor nodes. 
Their system can achieve 22 bit/s at a KDR of 2.2\% or 10 bit/s at a disagreement of 0.54\%. Ali~et al.~\cite{Ali2014Eliminating} investigated key generation in body area communications. The sensors were worn on the arm of a person and tested with scenarios including high activity (subject working and walking), low activity (subject seating and working) and dynamic environment (surrounding environment variation due to walking pedestrian). They used a Savitzky-Golay filter to improve the signal noise ratio (SNR) of the received signals. They demonstrated that it is feasible to leverage the existing communication packets for key generation, though it took about 15 to 35 minutes to generate a 128-bit key.

Recently, Li~et al.~\cite{li2017secret} proposed an RSSI trajectory-based secret key generation system for wearable devices. A bloom filter-based error correction scheme was proposed to correct the mismatches and Karhunen-
Loeve Transform (KLT)~\cite{dony2001karhunen} was used to enhance the randomness of the final key. The authors used both Universal Software Radio Peripheral (USRP) N210~\cite{usrp} and CC2530~\cite{instruments2012cc2530}. Evaluation results showed that their system is robust to eavesdropping attack and can generate a 128-bit key within 1s.

\paragraph{Wi-Fi}
Wi-Fi is one of the most successful wireless technologies, which has been widely installed in almost all laptops, tablets, smartphones, and many other consumable devices. 
Wi-Fi is based on the IEEE 802.11 technologies, hence we use Wi-Fi and IEEE 802.11 interchangeably in this paper.
Since IEEE 802.11 is first introduced in 1997, several amendments have been released, including IEEE 802.11a/b (1999), IEEE 802.11g (2003), IEEE 802.11n (2009) and IEEE 802.11ac (2013). Because there are numerous off-the-shelf Wi-Fi platforms, they have become the ideal testbed for key generation.

The work by Mathur~et al.~\cite{mathur2008radio} is probably the first Wi-Fi based key generation system. They used both the RSSI and the peak of the CIR. They designed a level-cross algorithm, which can achieve a high key agreement, hence information reconciliation is not required.
Their algorithm was evaluated both theoretically and experimentally.

However, since RSSI and the peak of CIR can only provide coarse-grained information about wireless channels, the key generation rate of previous systems is largely limited even with the help of multi-bit quantization. To break this barrier, researchers started to explore fine-grained channel information that can be extracted from OFDM. Luckily, researchers find that CSI contains rich information about multiple sub-carriers of OFDM. Especially the release of tools that can extract CSI from commodity Wi-Fi devices~\cite{halperin2011tool} open the door for CSI-based key generation systems which later shows significant performance improvement compared to its counterparts. In particular, Liu~et al.~\cite{liu2012exploiting} carried out the first theoretical analysis for CSI-based key generation when there is a multipath channel. Then, Liu et al.~\cite{liu2013fast} proposed the first practical CSI-based key generation system which achieved key generation rate of 60-90 bit/packet. Since then, numerous studies have been conducted to investigate CSI-based key generation systems from different application scenarios and perspectives~\cite{zhao2012efficient,xi2014keep,zhang2016efficient,zhang2019design,zhang2016experimental}. 

Zhang~et al.~\cite{zhang2016efficient} modelled and analysed the time and frequency correlation of the OFDM by using Wi-Fi as a case study.
Zhang~et al.~\cite{zhang2019design} took a step further by extending key generation to multiple users. In particular, they leveraged the Orthogonal Frequency-Division Multiplexing Access (OFDMA) for enabling the access point to communicate with multiple users simultaneously. However, Xi et al.~\cite{xi2014keep} found that CSI measurements from adjacent subcarriers have strong correlations and hence the generated keys may have low randomness. To solve this problem, they proposed a key generation scheme, named KEEP, which adopted a validation recombination mechanism to generate secret keys from CSI measurements of all subcarriers.



\paragraph{LoRa}
LoRa is a new IoT wireless technique, which features low power, low data rate and long range. The first LoRa-based key generation works were in 2018~\cite{ruotsalainen2019experimental,zhang2018channel,xu2019lora}. Compared to the short range communications-based key generation, LoRa-based key generation has yet received less attention. LoRa protocol only provides RSSI information.

Ruotsalainen~et al.~\cite{ruotsalainen2019experimental} investigated the effects of LoRa setup on the key generation performance. They carried out extensive experiments with different configurations of spreading factors and bandwidths. These parameters will affect the airtime of LoRa packets hence the sampling delay between bidirectional transmissions between Alice and Bob. They also applied key generation with LoRaWAN specification and carried out experiments in indoor and outdoor suburban environments. They demonstrated that their system can even work when there was no significant channel variation. Xu~et al.~\cite{xu2019lora} designed a complete LoRa key generation protocol. They carried out extensive experiments with static and mobile modes including walking, biking and driving in indoor and outdoor environments.
They employed several signal processing algorithms to improve the key generation rate. They also designed a compressive sensing-based reconciliation framework to reduce the KDR. Their experiment results demonstrated key generation rates of 18 bit/s and 31 bit/s in stationary and mobile scenarios, respectively. Zhang~et al.~\cite{zhang2018channel} designed a differential value-based key generation protocol. They found the received power varied significantly when the LoRa devices moved in a large scale environment. In this circumstance, the threshold-based quantizer would produce non-random keys. In contrast, they quantized keys based on the trend of the received power. They carried out experiments in an indoor building and outdoor urban environment. The experimental results validated the differential value-based protocol can produce random keys. 

\paragraph{Bluetooth} Although Bluetooth has been widely installed in smartphones and wearable devices, there are very few key generation explorations with Bluetooth and the first work is in 2014 by Premnath~et al.~\cite{premnath2014secret}. 
Bluetooth operates at the 2.4~GHz ISM band, which is shared with Wi-Fi, ZigBee and microwave, etc.
Bluetooth uses frequency hopping to find the free spectrum and avoid collision with other techniques. Premnath~et al.~\cite{premnath2014secret} adopted this feature and their system can still work even under heavy Wi-Fi traffics.

\paragraph{5G:} The fifth-generation (5G) will be available very soon and has adopted a number of new and advanced physical layer modulation techniques, e.g., massive Multiple-input and Multiple-output (MIMO), millimeterwave (mmWave) communications, full duplex. 
These techniques provide new approaches to exploit channel randomness more efficiently.
A tutorial on key generation with 5G can be found in~\cite{li2019physical}.

Jiao~et al.~\cite{jiao2018CNS} investigated key generation performance for a mmWave MIMO system, which exploited the virtual angle of arrival (AoA) and angle of departure (AoD) characteristics.
A new channel estimation method was proposed to exploit the sparsity of the mmWave MIMO channel. Their simulation achieved above 99\% bit agreement ratio under very low SNR (-10dB), which can significantly decrease the reconciliation cost. Chen~et al.~\cite{chen2020beam} proposed a pilot reused key generation protocol for multi-user massive MIMO systems. Specifically, they designed two algorithms, namely beam domain-based channel probing and interference neutralization-based multi-user beam allocation, in order to reduce the channel dimension and reuse pilots. Their simulation results demonstrated a significant reduction of the channel estimation overhead. Vogt~et al.~\cite{vogt2018secret} employed full duplex communications for the channel probing to improve key generation performances. Full duplex can decrease the sampling timing and significantly improve the key generation rate. However, the residual self-interference of the full duplex will impact channel measurements~\cite{vogt2016practical}. The authors also found that full duplex channel probing can downgrade the eavesdropping attack because of the superposition of probing signals.

\subsubsection{Proximity-based Approaches}
All the systems described in Section~\ref{sec:reciprocity_radio} depend on channel reciprocity to generate keys. There is another research direction that exploits the co-location property of mobile devices to generate keys. 
As shown in Fig.~\ref{fig:model}(b), when two devices are physically co-located, their radio signals will be very close to each other. Below, we review several representative works using the co-location property of IoT devices.

Amigo~\cite{varshavsky2007amigo} is one of the earliest studies that use a common radio environment to authenticate mobile devices without explicit user involvement. It adopted Diffie-Hellman protocol (also known as D-H protocol) to establish keys followed by a commitment scheme to address Man-in-the-Middle (MITM) attack. Finally, the similarity between the radio signals measured by two devices is used to verify if they are in close proximity. The authors claimed that Amigo is secure against MITM attack, eavesdropping attack and spoofing attack.

Mathur et al.~\cite{mathur2011proximate} proposed Proximate, a system that generated a secret key for two mobile devices nearby from their measured wireless radio signal. Proximate adopted the conventional key generation process mentioned in the last subsection: quantization, reconciliation and privacy amplification. It removed the reliance on Diffie-Hellman protocol, but its bit generation was very low (only 1-3.5 bit/s).

To overcome the low bit rate problem, Xi et al.~\cite{xi2016instant} proposed a CSI-based authentication and key generation system called The Dancing Signals (TDS) for two devices in close proximity. Although TDS used CSI to generate keys, it presented a novel key generation method. The keys were generated randomly and encoded by the CSI features. Only the other device that had similar features can decode the key. Therefore, both devices can agree on the same key. By not generating keys from CSI directly, TDS improved key generation rate significantly. According to their evaluation, TDS can achieve a bit rate of hundreds of bit/s. Moreover, TDS can be extended to support a group of users.

Some other proximity-based schemes are based on another observation. That is, if a nearby sender moves very close to one of the antennas on the receiver, the receiver can observe a large RSSI variation. Instead, if a faraway sender moves close to the receiver, the two antennas will not see a large RSSI difference. Some representative works are Neighbor~\cite{cai2011good}, Wanda~\cite{pierson2016wanda} and Move2Auth~\cite{zhang2017proximity}. Good Neighbor is the first device pairing scheme based on this idea. Wanda was built on Good Neighbor but expanded to generate secret keys based on the channel reciprocity. Move2Auth borrowed the idea from Good Neighbor and Wanda and used for a smartphone to authenticate a nearby IoT device.

Co-location based approaches take advantage of the physical proximity to authenticate devices. However, these approaches suffer from a common problem: the distance between two legitimate devices should be close, e.g., 1.25 cm in Proximate~\cite{mathur2011proximate}, 5 cm in TDS~\cite{xi2016instant} and 20 cm in Move2Auth~\cite{zhang2017proximity}. Therefore, the practicability of such approaches is low because the wireless transceivers are embedded in mobile devices and in some scenarios it is hard to put two antennas in such short distance.

\subsection{Audio}
\label{subsec:audio}
Currently, microphones and speakers are integrated in many IoT devices such as smartphone, Google assistant, Amazon Echo and laptops. Acoustic waves, as a form of wave, possess many similar properties as radio waves such as fading, multi-path, reflection and diffraction.  Accordingly, a large portion of existing work have studied how to use audio signals to pair IoT devices. We describe several representative systems that achieve device pairing by audio signals. Same as radio-based key generation, the majority of works can be divided into two classes: channel reciprocity-based and proximity-based.

\subsubsection{Channel Reciprocity-based Approaches}
Researchers also demonstrated that acoustic channel holds reciprocity~\cite{lu2019free}. Therefore, several acoustic channel reciprocity-based device pairing systems have been proposed recently. 

Lu et al.~\cite{lu2019free} conducted the first study to verify the reciprocity of acoustic channel. 
In order to not disturb users, FREE used the inaudible frequency range 18k-22kHz. FREE works as follows. To estimate the acoustic channel, FREE transmits a pre-defined sequence, which is modulated to the symbols of Gaussian Filtered Minimum Shift Keying (GMSK). The generated signal is stored in a Waveform Audio (WAV) file and then transmitted by the speaker. After two devices exchange a number of messages, they use channel taps~\cite{tse2005fundamentals} as acoustic channel features to generate keys. The authors implemented FREE on several smartphones and evaluated its performance in four environments: indoor static environment, indoor mobile environment, outdoor static environment and outdoor mobile environment. The results showed that FREE worked well when two devices were within 60 cm distance. They validated the randomness of the generated keys by National Institute of Standards and Technology (NIST) test and also analysed the security of FREE against possible attacks. 


Bala et al.~\cite{bala2020phy} also proposed a key agreement system based on acoustic channel reciprocity. The proposed system adopted the conventional key generation procedure in wireless key generation: channel probing, quantization, reconciliation and privacy amplification. Different from FREE, the proposed system used sound pressure levels as acoustic channel physical layer characteristics. The authors implemented the proposed system on Samsung Galaxy On5 pro smartphone and conducted evaluation in different environments including classroom, conference room and hallway. By carefully tuning parameters, they can achieve a key generation rate of 80 bits/sec and a bit error rate of $25\%$ before reconciliation. They also conducted NIST test~\cite{rukhin2001statistical} and the generated keys pass the test.  


\subsubsection{Proximity-based Approaches}
As shown in Fig.~\ref{fig:proximitybased}, the idea of proximity-based scheme is that sound recorded by microphones do not vary too much within close proximity, but significantly differ when two devices are far away from each other.

Sch{\"u}rmann and Sigg~\cite{schurmann2011secure} proposed a system to establish a common cryptographic key for devices in the same context based on ambient audio patterns. The system works as follow. First, both devices need to perform a Network Time Protocol (NTP)-based synchronisation method to establish sufficient synchronisation. This is because synchronisation among devices plays a significant role in their approach. The extracted fingerprints will be totally different if the start time of audio signal differs several hundreds milliseconds. Second, both devices extract a binary representation of their recorded audio which is called audio fingerprint. The binary sequences of two devices will be highly similar to each other if they are in close proximity. Finally, they use fuzzy commitment scheme to account for the errors between the binary sequences to generate the same key. 

The authors verified the feasibility of the proposed approach in several realistic environments including  office environment, canteen environment and outdoor environment. The results showed that their protocol works well in an office environment. In the crowded canteen environment, the matching rate will decrease if there is not a dominant audio source. In the outdoor environment (a heavily trafficked road), it is hard to establish a secure communication channel based on ambient audio only. They also analysed the security of the proposed pairing scheme against various potential attacks such as eavesdropping, brute force attack, MITM, Denial-of-Service (DOS) and audio amplification attack. The analysis showed that these threats can be eliminated by a careful choice of the fingerprint mechanism.

Following the same idea, Karapanos et al.~\cite{karapanos2015sound} proposed Sound-proof, a two-factor authentication scheme based on ambient sound. The idea of Sound-proof is that if a smartphone is close to a computer, they will record similar sound signals. So the similar sound signals can be used as the second factor to verify user's identity. The authors implemented Sound-proof on many platforms. The browsers include Google Chrome, Mozilla Firefox, and Opera. The smartphone platforms include Samsung, Google Nexus, Sony, Motorola, and different iPhone models. The evaluation in both indoor and outdoor environments showed that Sound-proof can achieve an Equal Error Rate (EER) of 0.002. EER is the crossing point of false rejection rate and false acceptance rate. It means, out of 100 trials, only $0.2\%$ of genuine user's attempts are rejected and $0.2\%$ of attacker's attempts are accepted. Sound-proof can complete verification within 5 seconds and work well at different locations and distances.


The primary limitation of the above two systems is that both devices need an accurate synchronisation which is sometimes unrealistic especially for devices that meet each other for the first time. AudioKey proposed by Shang and Wu~\cite{shangaudiokey} aims to pair two smart watches when two users shake their hands together. It adopts the same idea as the above work but eliminates the requirement of synchronisation. The working flow of AudioKey is as follows. The pairing process is triggered by detecting the fist negative peak during the handshake process. The observation is the acceleration signals measured by two smart watches should follow the same pattern and reach peak almost at the same time because two hands are held together tightly. AudioKey extracts binary key sequence from both time domain features and frequency domain features. The Golay code G(24, 12) is then employed to correct the errors between two smartwatches.

The authors recruited nine volunteers whose ages were from 22 to 29 to evaluate the performance of AudioKey. The results showed that it can achieve a bit generation rate of 13.4 bit/s and key agreement rate of $96.7\%$ for a 128-bit key. They also analysed the security against mimicking attacker. If an attacker located 1.2 m away started key generation at the same time as the genuine user, his matching rate can be as high as $73.25\%$. Although the authors claim that the location of the mismatched bits are unknown to the attacker, the entropy of the key is decreased significantly, and the security needs further validation.

Genewave proposed by Xie et al~\cite{xie2017genewave,xie2018genewave} is also based on acoustic signal proximity. Their idea is that the acoustic channel response is unique for a given device at a given location. The proposed system includes two major steps: bidirectional initial authentication and key agreement. In the first step, they use the round-trip time of acoustic signal between two devices to authenticate each other. Specifically, if the response interval is larger than a pre-defined threshold, the device will be regarded as an attacker. In the second step, Alice and Bob first build the features of the acoustic channels. Then, the bits `1' and `0' are encoded in the acoustic signal based on a novel sine wave-based pulse coding method. Suppose Alice encodes bit sequence in an acoustic signal and transmit the modulated signal to Bob. Bob can decode the same symmetric key based on the acoustic channel response between Alice and Bob. So both Alice and Bob establish the same key to secure their communication.

The authors implemented GeneWave on Nexus smartphone and conducted evaluation in three environments: line-of-sight (LOS) meeting room, non line-of-sight (NLOS) meeting room and coffee shop. However, they cleared all human activity in meeting room to mitigate the influence of environment changes. The evaluation results showed the proposed system can achieve high matching rate in the above three environments. Moreover, GeneWave can complete authentication and generate a 2048 bits key in 2 s, which is 10$\times$ faster than TDS~\cite{xi2016instant}. Although the evaluation showed GeneWave can pair two devices quickly, the evaluation is conducted in a well controlled environment. First, Alice and Bob need to be very close to each other (<3cm) to achieve high matching rate. Second, there should be no activities in the nearby environment to avoid multi-path changes. Therefore, the usability of GeneWave in a realistic environment still requires further study.

Although the systems above can achieve high performance as demonstrated in the evaluation, they can only achieve D2D pairing. It is desirable that a group of IoT devices can be paired together based on the sound if there are a number of devices present. Motivated by this, Gu et al.~\cite{gu2016scalable} proposed a group audio-based authentication scheme for IoT devices which is called GAB-IoT. In GAB-IoT, there is a central device who broadcasts audio signal to nearby IoT devices. All the devices in an audio-reachable distance can observe similar audio signal, based on which they can generate the same key using fuzzy extractor~\cite{dodis2008fuzzy}.
\subsection{IMU Sensors}
In this subsection, we review representative device pairing systems based on sensors embedded in IoT devices. The on-board sensors in IoT devices provide them capabilities to collect information about the ambient environment. Thus the sensory information can be used to complement the limitation that two devices have no pre-shared secret. So far, a variety of sensory signals have been used as common knowledge to assist IoT device authentication and key generation. In the following, we survey this type of work based on their sensor types.
  
A large number of device pairing systems have been proposed by utilising the built-in IMU sensors. IMU sensors include 3-axis accelerometer, 3-axis gyroscope and 3-axis magnetometer. Among these three types of sensors, accelerometer is the most widely used sensor to design a device pairing system. This is because it provides a good source of entropy for bootstrapping a secure communication channel in autonomous and spontaneous interactions between mobile devices that share a common context but were not previously associated. The shared common context is usually a daily activity such as shaking, walking, hand gesture etc. Next, we discuss this category of work based on the type of motion used in each scheme.
\begin{figure*}[!t]
	\centering
	\hspace{-0.1in}
	\subfigure[Shake well before use~\cite{mayrhofer2009shake}]{
		\includegraphics[height=1.2in]{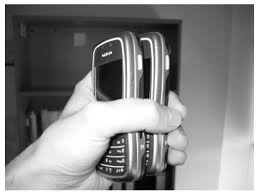}
		\label{fig:shake-well-before-use}}
		\subfigure[Walkie-Talkie~\cite{xu2016walkie}]{
		\includegraphics[height=1.2in]{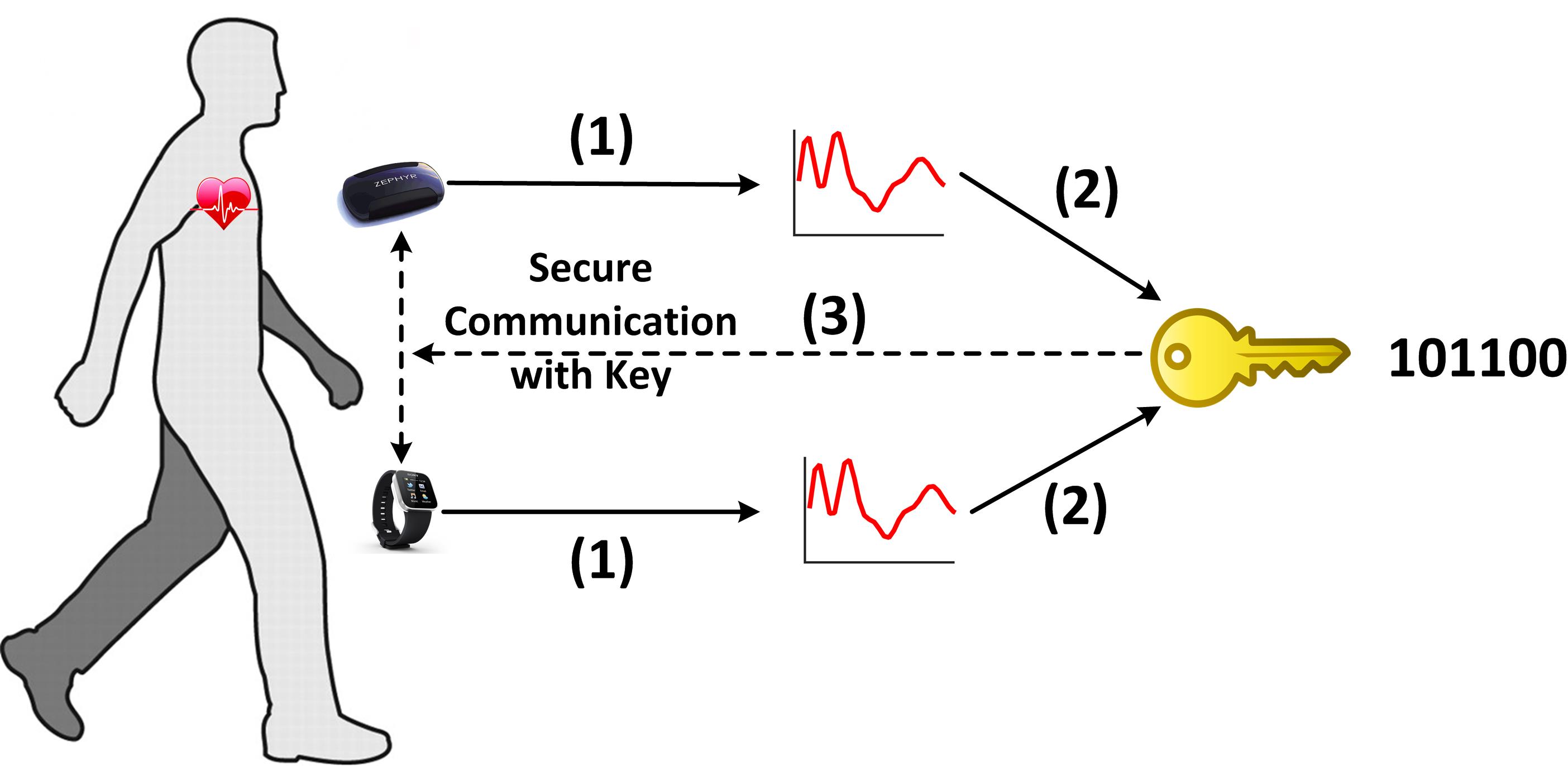}
		\label{fig:walkie-talkie}}
		\hspace{-0.1in}
		\subfigure[Shake-n-Shack~\cite{shen2018shake}]{
		\includegraphics[height=1.2in]{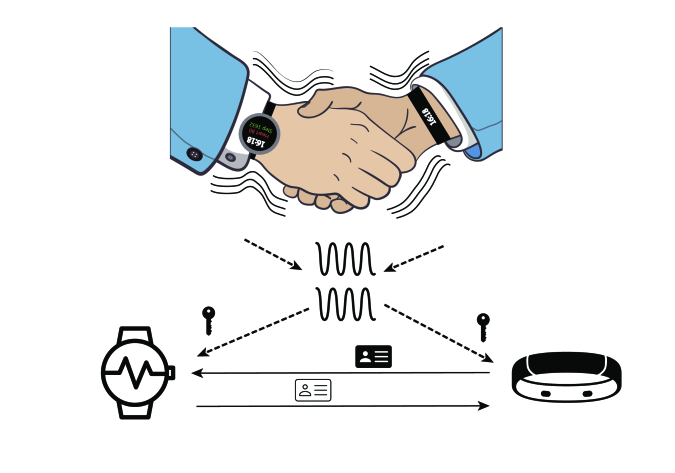}
		\label{fig:shake-n-shack}}
		\hspace{-0.1in}
	\caption{Representative works of accelerometer-based device pairing system.}
	\label{fig:accelerometerbasedscheme}
	\vspace{-0.2in}
\end{figure*}

\subsubsection{Shaking-based scheme}
The idea of shaking two devices together to pair them was first proposed by Holmquist et al.~\cite{holmquist2001smart}. To verify this idea, the authors implemented a prototype called \textit{Smart-Its Friends}, which is a small embedded artefact that can be associated when a user holds and shakes them together. \textit{Smart-Its Friends} pairs two devices without explicitly generating keys. Since then, a large body of studies based on this idea have emerged. 

The Martini Synch proposed by Kirovski et al.~\cite{kirovski2007martini} is the first protocol that generates a common key between two accelerometer-equipped devices shaking together. It adopted a simple fuzzy quantizer and an off-the-shelf cryptographically secure hash function, which they called joint fuzzy hashing protocol. By implementing a prototype using off-the-shelf components, they showed that their method can produce between 9–20 bits of entropy per second.

Mayrhofer and Gellersen~\cite{mayrhofer2009shake} designed the first complete pairing protocol, Shake well before use. Specifically, they proposed two schemes ShaVe and ShaCK. ShaVe first used Diffie-Hellman protocol to establish an insecure channel. It is known that standard Diffie-Hellman protocol is susceptible to MITM attack. So the authord used the Interlock protocol~\cite{rivest1984expose} for protecting against MITM to exchange the acceleration data. Finally, two series of acceleration data measured by two devices were compared to each other based on coherence. If it passes check (the coherence exceed a predefined threshold), the key generated by D-H protocol is used as session key to secure their subsequent communications. ShaCK adopted a direct key generation method. First, the FFT features were calculated from the captured acceleration data. Then, the FFT features were quantized to binary keys based on candidate key protocol which accumulated 
matched parts of the keys until it reached sufficient entropy.

In addition to the systems above, there are also a number of key generation systems based on shaking motion, e.g., Shake on it (Shot)~\cite{studer2011don}, Shakeunlock~\cite{findling2014shakeunlock}, ShakeMe~\cite{yuzuguzel2015shakeme}, and iShake~\cite{shi2019ishake}. They essentially use the same idea but different methods to generate keys. For example, Shakeunlock uses coherence similarity to determine whether two devices are shaking together while ShakeMe extracts discriminative features from the accelerometer data to derive keys.

Recently, another daily activity, handshaking, has also attracted researchers' interest. Handshaking is a common practice when two persons meet to express trust and respect. The motion pattern during this process can be captured by the motion sensors in user's wrist-worn wearables and then employed to extract secret keys for establishing a secure communication channel. The first idea was proposed and demonstrated in Shake-n-Shack~\cite{shen2018shake}. The evaluation results showed that Shake-n-Shack can generate 128-bit keys around 1 s with success rate >$99\%$. Motivated by Shake-n-Shack, Jiang et al.~\cite{jiang2019shake} proposed another acceleration-based pairing scheme for wrist-worn devices. They enhanced the efficiency of Shake-n-Shack by using an improved fuzzy cryptography scheme~\cite{anees2018discriminative}. But their result was obtained from Matlab simulation in a laptop rather than implementing the system on real wrist-worn devices. 

Shaking is an intuitive movement with natural variance. Additionally, it is easy to learn and perform. However, the usability of shaking-based approaches is restricted by the shape, size, and weight of the involved devices~\cite{chong2010classification,chong2012usability}.

\subsubsection{Walking-based scheme}
Shaking devices is intuitive, vigorous, and varying but only applies to wearable devices that can be held in the hand or worn on the wrist. In recent years, researchers have found another common daily activity that can be used to generate keys, i.e., gait. Gait refers to people's walking patterns. Like other biometrics such as face and fingerprint, gait is a biometric characteristic and studies from psychology and biometrics have shown that different people have distinguished and unique walking patterns~\cite{zhang2004human,ren2013smartphone,xu2017gait}. Therefore, mobile devices worn on the same user's body can measure the same gait pattern while devices on other user's body have different measurements. This fact lays the foundation for walking-based key generation schemes. 

Lester et al.~\cite{lester2004you} first proposed to use low-cost accelerometer to determine whether two devices are carried by the same person or not by analysing the coherence of the walking data. Their method worked well when two devices were carried in the same location on the body. However, there are two limitations in their study. First, there were only two subjects involved in the experiment so the results did not fully demonstrate the feasibility. Second, their method only achieved approximately $87\%$ accuracy if two devices were carried at different locations on the body.

Later, Cornelius and Kotz~\cite{cornelius2012recognizing} extended the work of Lester et al. to sensors carried at different locations on the body including wrist, ankle, and waist. They first extracted seven features that are commonly used in activity classification. Then they employed supervised machine learning method, namely support
vector machine (SVM), to determine whether two devices were carried on the same body. They used a dataset of seven subjects walking for 22 minutes to evaluate their approach and showed that their approach outperformed the work of Lester et al.~\cite{lester2004you} when devices were on the different locations on the body.

Motivated by these two pioneering works, Xu et al.~\cite{xu2016walkie} proposed the first gait-based key generation protocol for on-body mobile devices Walkie-Talkie. It followed the traditional key generation process as in radio-based key generation introduced in Section~\ref{sec:radio}: quantization, reconciliation and privacy amplification. It can generate a 128-bit key within 5 seconds ($\approx$26 bit/s). The performance was further improved in their extension paper Gait-key~\cite{xu2017gait}. Gait-key utilised a multi-bit quantization approach to improve key generation rate and further employed error-correction code to correct the mismatch between the initial keys. The time required to generate a 128-bit key was reduced from 5 seconds to 3.5 seconds ($\approx$ 36.6 bit/s). The authors also implemented Walkie-Talkie and Gait-key on MoTo E2 smartphone to evaluate the system cost. Their results showed that the computation time of Walkie-Talkie and Gait-key is 316.8 ms and 419.2 ms, respectively. They analysed the security of Walkie-Talkie and Gait-key by asking attackers to mimick genuine user's walking patterns. They found that a mimicking attack can at most achieve about $50\%$ agreement rate indicating that their systems are resilient to active mimicking attack.

In another work BANDANA~\cite{schurmann2017bandana}, the authors also used gait to authenticate devices on the same body. Different from Walkie-Talkie, BANDANA proposed a novel quantization method and utilised fuzzy commitment scheme~\cite{juels1999fuzzy} to account for the errors. Compared to the fuzzy vault scheme, fuzzy commitment is less complex and computationally expensive in terms of key locking and unlocking. Their evaluation results showed that BANDANA can achieve $80\%$ similarity between devices on the same body. Because they only used the acceleration data along gravity direction to generate keys, the key generation rate was very slow: it took approximately 96 s to generate a 128 bit key ($\approx$ 1.3 bit/s). 

The authors in~\cite{sun2017secure} followed the same idea and simply combined the advantage of Walkie-Talkie and BANDANA together. Their method used 3-axis acceleration data to generate keys and employed fuzzy commitment scheme with BCH code to correct errors. However, their key agreement rate can only reach $79\%$.

In a more recent work Auto-Key~\cite{wu2020auto}, Wu et al. proposed to use autoencoder to speed up the key generation rate of gait-based scheme. Instead of exchanging additional error correction information, one device uses an autoencoder to predict the gait measurements at another device attached to the same body. Then the key was generated based on the predicted accelerometer data. The evaluation results showed that Auto-Key can improve matching rate by $16.5\%$ while speeding up bit rate by 1.9$\times$.

However, the systems mentioned above suffer the same limitation: they can only pair two devices and fail to pair a group of devices. To solve this problem, Revadigar et al.~\cite{ revadigar2017accelerometer} proposed a group key generation scheme for wearable devices based on user's walking patterns. The system works as follow. First, a hub device generates a random key using random number generator (RNG). Then, it uses fuzzy vault scheme~\cite{juels2006fuzzy} to construct a vault based on the measured acceleration data. Afterwards, it broadcasts this vault to nearby devices. The other devices can recover the random key by unlocking the vault using their own gait measurements. According to fuzzy vault scheme, only highly similar gait signal can be used to unlock the vault. Therefore, their approach ensures that the random key hidden in the vault can only be recovered by the devices on the same user's body. Because the gait signal is not used to generate keys directly, the key generation rate is up to 750 bit/s.

Since walking is a daily activity, walking-based key generation systems provide an autonomous and spontaneous way to pair wearable devices that are worn on the same user's body. Although all the authors above claimed that their approaches can generate keys with high randomness, a recent study revealed that there is a bias in the generated binary bit strings~\cite{bruesch2019security}. They also pointed out that the attacker can generate a highly similar key by video analysis. Unfortunately, there has not been any work in the literature that can prevent this kind of attack. Therefore, it would be an interesting future research topic.

\subsubsection{Others}
Apart from the motions above, other motions are also utilised to facilitate secure device pairing. 
Synchronous gestures such as bumping was utilised to pair two mobile devices equipped with accelerometers. The intuition is bumping generates equal and opposite hard contact forces that are simultaneously sensed
by the accelerometer. Hence, the accelerometer signal can provide enough information to determine whether two devices are physically interacting or not. Example of such systems include Hinckley et al.~\cite{hinckley2003synchronous} and BUMP~\cite{BUMP}.

Wang et al.~\cite{wang2016touch} introduced Touch-and-Guard (TAG), a system that used hand resonant to associate a wrist-worn wearable with an external device equipped with accelerometers. TAG is based on the observation that the hand and the external device form a vibration system of which resonant properties measured by two devices attached to hand or wrist are highly correlated. The authors demonstrated the feasibility of TAG by implementing a prototype on Arduino development board and conducted experiments involving 12 subjects. Evaluation results showed that TAG can achieve a key generation rate of 7.84 bit/s, which is faster than commonly used PIN authentication scheme. They also analysed the security of TAG and found that it is resilient to passive acoustic eavesdropper but vulnerable to visual eavesdropper using high-speed cameras.

Han et al.~\cite{han2017convoy} proposed to use road bumpiness and traffic conditions as a common context to detect trucks in the same platoon. Convoy (the name of their system) exploited the fact that trucks in the same platoon experience similar road (e.g., bumps and cracks) and traffic (e.g., acceleration and steering) conditions. By using the embedded accelerometers, Convoy used the traditional fuzzy commitment scheme~\cite{juels1999fuzzy} to establish the same key for the trucks in the same platoon. The authors implemented and tested the Convoy protocol with real-world driving data. Evaluation results demonstrated that vehicles moving in neighbouring lanes can be differentiated adequately by their context, and Convoy can counter platoon ghost attacks.

\subsubsection{A Summary of Motion-based Schemes}
In this subsection, we have discussed various motion-based device pairing schemes. These methods require users to perform an explicit action (i.e., shake devices together). Because it explicitly involves users, it increases their perception of the association while it executes. The motions researchers have explored are common daily activities such as shaking, walking, bump and touch. Therefore, these schemes are intuitive and easy to learn. Nevertheless, their usability is limited by the shape and form factor of the involved IoT devices~\cite{chong2012usability}.

\subsection{Miscellaneous Hardware}
Advances in microelectronics, embedded system and wireless technology, have led to a rapid development of new types of miniature sensors in IoT devices which in turn create many new opportunities for context-based IoT device pairing. In this subsection, we survey recent related work using new hardware. It should be noted that due to the diversity of miscellaneous hardware, we categorise these works based on the signal they used to pair devices.

\subsubsection{Heartbeat Signal}
The concept of using human biometrics to secure body area network communication was first proposed by Cherukuri et al.~\cite{cherukuri2003biosec}. Conventionally, biometrics is a technique commonly known as the automatic identification or verification of an individual by his or her physiological or behavioural characteristics~\cite{jain2004introduction}. The concept described here, however, is different from conventional biometrics cryptography systems. In the context of key generation, biometrics traits are used as keying materials to generate random keys. To this end, the biometrics should fulfil the following two requirements as noted in~\cite{poon2006novel}.

\begin{itemize}
    \item Distinctiveness. The trait should be sufficiently distinctive on two individuals, so that the keys generated from different subjects will be different.
    
    \item Time variation and invulnerability. The trait should change over time and have a high degree of randomness so that the biometric characteristics recorded at different times would not generate the same key even they are obtained from the same person. 
\end{itemize}

\begin{figure*}[!t]
	\centering
	\subfigure[IMDGuard~\cite{xu2011imdguard}]{
		\includegraphics[height=1.3in]{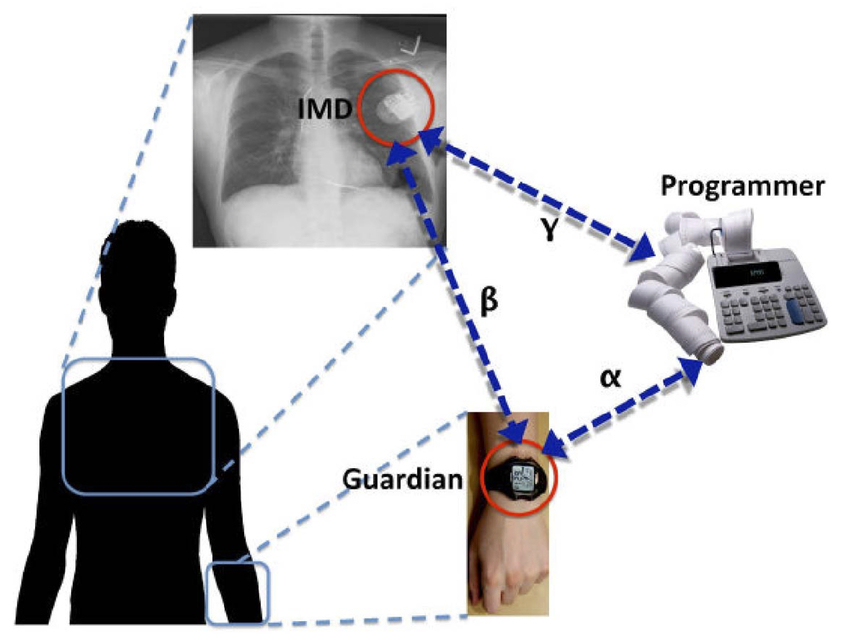}
		\label{fig:IMDGuard}}
		\subfigure[H2H~\cite{xu2016walkie}]{
		\includegraphics[height=1.3in]{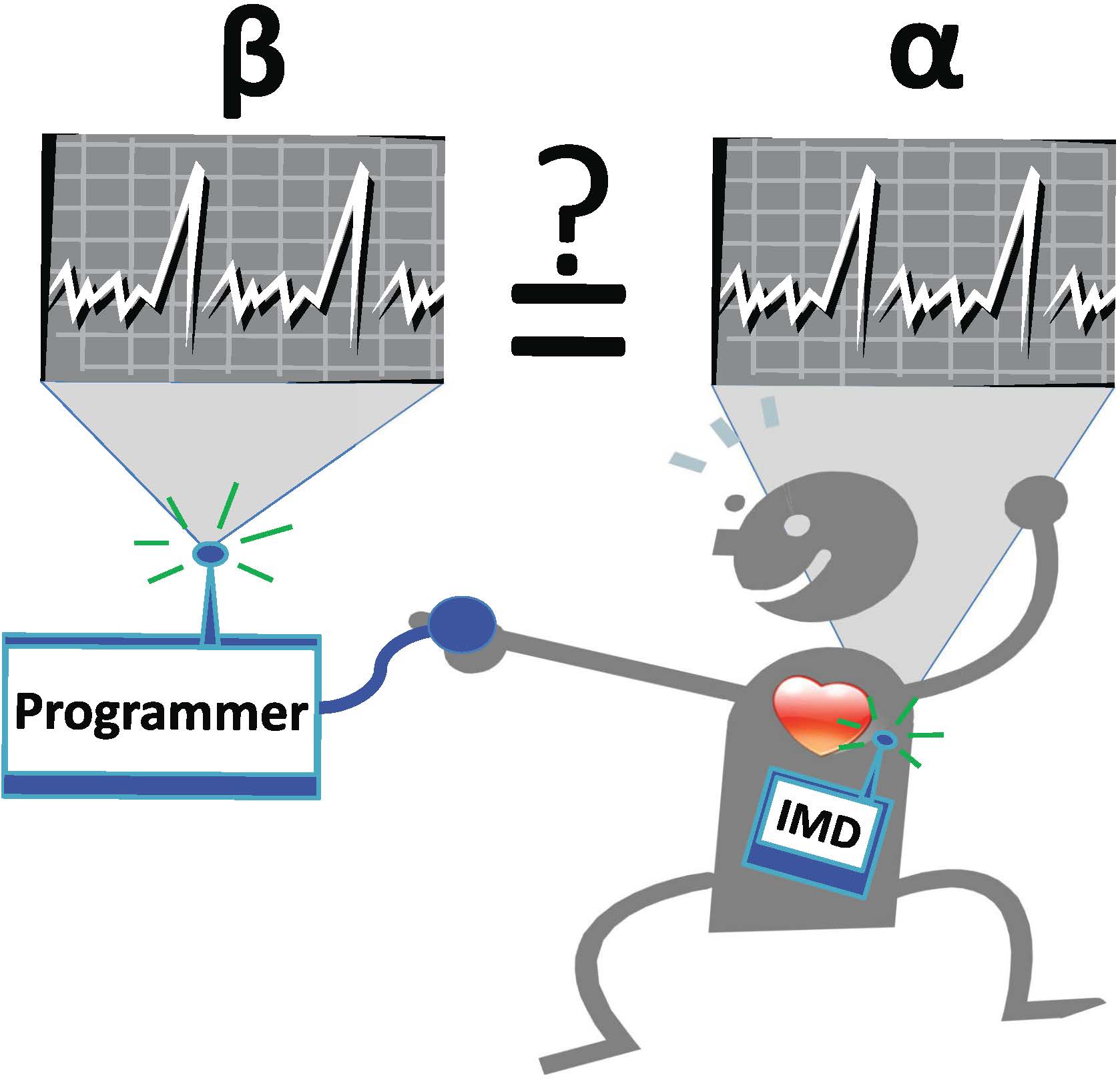}
		\label{fig:H2H}}
		\subfigure[H2B~\cite{lin2019h2b}]{
		\includegraphics[height=1.3in]{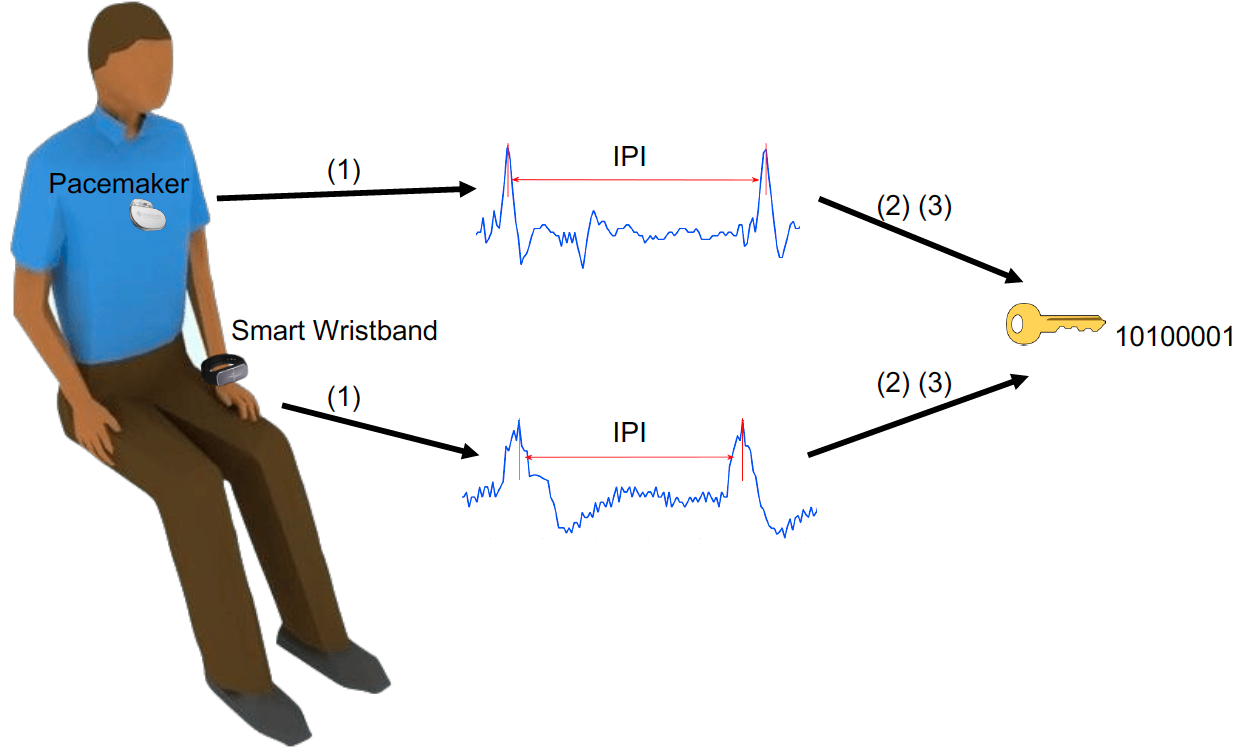}
		\label{fig:H2B}}
	\caption{Representative works of IPI-based key generation systems.}
	\label{fig:IPI}
	\vspace{-0.2in}
\end{figure*}

The motion-based systems described in the last subsection are essentially based on behavioural characteristics. Despite their user friendliness, they still pose some burdens on users. A desirable device pairing system should reduce user involvement as much as possible. Moreover, the feasibility of such behavioural-based device pairing system is validated on a small-scale dataset under controlled environment. It is still elusive whether they apply to large-scale population. However, another human physiological biometrics, heartbeat, does not have such problem because they have been validated on large datasets~\cite{cherukuri2003biosec,obrist2012cardiovascular}. Specifically, researchers have found that inter-pulse interval (IPI), i.e., the interval between heartbeat signal peaks, is highly random and can be used as a random source to generate keys~\cite{rostami2013heart}. Fig.~\ref{fig:IPI} illustrates the idea of heartbeat signal-based key generation system. Below, we discuss several representative works utilising IPIs to pair wearable devices.

The first practical key generation system based on IPIs was proposed by Poon et al. in their pioneering work~\cite{poon2006novel}. The suggested system was validated on a dataset of 99 subjects, and the results showed that it can achieve a total error rate of $2.58\%$ when the IPIs were coded into 128-bit key. Later, the same group of authors proposed a more advanced system which improved the system performance significantly~\cite{bao2008using}. The proposed system works as follows. To begin with, a series of IPIs were captured by biosensors located at different body locations. Then, multiple IPIs were accumulated to achieve better performance followed by a modulo operation to randomise the monotonically increasing multi-IPIs. Finally, the mapped IPIs were represented by their corresponding Gray code to form the final binary sequence. Compared to their previous scheme~\cite{poon2006novel}, the error rate with 64-bit key generated was reduced from $6.98\%$ is $2.83\%$. Meanwhile, the improved system required less IPIs to generate the same length of key. In both systems, the IPIs were obtained from biosensor signals that can record electrocardiogram (ECG) and photoplethysmogram (PPG).

However, Xu et al.~\cite{xu2011imdguard} found that previous work did not properly extract the randomness of IPI. Specifically, as the average IPI of most people is about 850 ms, the 7th and 8th digits of the ending time are not random at all. For example, suppose the first IPI value is 860 ms and its Gray code representation is ''\underline{10111}10010''. If there is another IPI whose value is 840 ms, then its Gray code representation is ''\underline{10111}01100''. It is clear that the first 5 digits are exactly the same. The randomness of IPI data lies in the least significant bits, so does the error. To tackle this problem, Xu et al.~\cite{xu2011imdguard} proposed the first comprehensive secure protocol for ECG-based key agreement system, IMDGuard. IMDGuard solved this issue by proposing an algorithm that can transform raw IPI data from normal distribution to uniform distribution. Then, the least 4 bits were used to generate keys. Finally, IMDGuard used parity check to discard the mismatched bit blocks to generate the final key. The authors implemented IMDGuard on TelosB board which utilised the CC2420 transceiver for wireless communication. Evaluation results showed that IMDGuard required 61 IPIs, corresponding to 45 seconds or so, to generate a 128-bit key. The author also validated the randomness of the extracted keys by NIST test suite~\cite{rukhin2001statistical}.

Venkatasubramanian et al.~\cite{venkatasubramanian2008plethysmogram,venkatasubramanian2009pska} proposed to use fuzzy vault to account for the measurement errors between different biosensors. The difference between these two works is in~\cite{venkatasubramanian2008plethysmogram} they used PPG only while in~\cite{venkatasubramanian2009pska} they utilised the combination of PPG and electrocardiogram (EKG). However, the security level of these two systems is not high due to the limited feature size and high computational overhead. To overcome this limitation, Hu et al.~\cite{hu2013opfka} proposed IPI-based key agreement protocol termed OPFKA. In comparison with the work of Venkatasubramanian et al.~\cite{venkatasubramanian2008plethysmogram,venkatasubramanian2009pska}, OPFKA is more secure and energy efficient. OPFKA is based on the observation that secret features generated by a sensor are ordered and only the sensor itself is aware of the order of the features. But this way is controversial because the authors assume only Alice and Bob know the order of features which is actually a pre-shared secret. It contradicts the normal assumption that two devices meet for the first time have no pre-shared secret between them.

However, as discussed in~\cite{rostami2013balancing}, IMDGuard and OPFKA have serious security flaws. For instance, by performing a simple MITM attack to IMDGuard, the effective key length is reduced from 189 bits to 86 bits. To solve this problem, Rostami et al.~\cite{rostami2013heart} proposed their ECG-based authentication protocol, Heart-to-Heart (H2H). H2H is a novel cryptographic device pairing protocol that uses the randomness of IPI to protect against active attackers, while meeting the requirement of lightweight implementation and noise tolerance in ECG readings. The authors implemented H2H in an ARM-Cortex M-3 microcontroller to demonstrate the practicality of H2H in current implantable medical device (IMD) hardware. The authors claimed that H2H is the first physiologically-based IMD device pairing protocol with a rigorous adversarial model and protocol analysis.

With the emergence of new sensors, researchers started to investigate the feasibility of using these sensors to detect heartbeat signals. Lin et al.~\cite{lin2019h2b} proposed the first piezoelectric sensor-based key generation system H2B for on-body wearable and implantable devices. H2B is based on the observation that the minor vibrations caused by heartbeats can be detected by piezoelectric sensors on different body locations. The evaluation results showed that H2B can pair two wearable devices on the same user’s body with a success rate of $95.6\%$. Security analysis of H2B also demonstrated that it is secure against passive attack, active presentation attack and active video attack.

Compared to walking and shaking, heartbeat signal-based key generation systems can ease the burden on users, i.e., users do not need to walk a few steps or shake devices. Additionally, several large dataset are publicly available online. For example, the dataset used in~\cite{venkatasubramanian2008plethysmogram,venkatasubramanian2009pska,xu2011imdguard,hu2013opfka} is PhysioBank
database \footnote{http://www.physionet.org/physiobank}, and the dataset used in~\cite{rostami2013heart} includes MIT-BIH Arrhythmia Database~\cite{moody2001impact}, PTB Database~\cite{bousseljot1995nutzung}, and MGH/MF Waveform Database~\cite{welch1991massachusetts}. However, due to the limited entropy of information source, this kind of scheme cannot achieve high key generation rate. For instance, the state-of-the-art heartbeat signal-based system H2B can generate keys at the speed of 3 bit/s, which is several order-of-magnitude lower than walking-based scheme (750 bit/s in Revadigar et al.~\cite{ revadigar2017accelerometer} ) and shaking-based scheme (128 bit/s in Shake-n-Shack~\cite{shen2018shake}).

\subsubsection{Gait Signal} 
The aforementioned walking-based device pairing schemes use accelerometer to record user's walking patterns. However, the main problem of accelerometer-based sensing systems is continuously sampling accelerometer will quickly drain the battery of resource-constrained IoT devices. To address this problem, a recent trend is to use kinetic energy harvesting (KEH) device to replace accelerometer. KEH is the process of converting motion and vibrations into electrical energy to power electronic devices. Human activities are the most relevant sources because wearable devices can harvest energy directly from user activities. Therefore, KEH is a promising technology for applications where batteries are impractical, such as wearable and implantable sensors~\cite{kiziroglou2012materials}.

Initially, this idea is proposed in the field of activity~\cite{khalifa2017harke}, gait~\cite{xu2017keh,xu2018keh} and transport mode classification~\cite{lan2019entrans,lan2016transportation} for the purpose of power saving by not sampling accelerometer. The first piezoelectric energy harvesting-based key generation system based on user's walking pattern was proposed by Qi et al.~\cite{lin2020kehkey} until very recently. The principle of KEHKey (the name of their system) is the same as walking-based key generation schemes such as Walkie-Talkie~\cite{xu2016walkie}, i.e., the energy patterns harvested by multiple KEH devices on the same user's body are highly correlated while it is not accessible to outsiders.
 
KEHKey consists of the following three stages: sampling and smoothing, quantization, and information reconciliation. First, Alice and Bob harvest energy from user's walking independently. Then, they quantize the measurements into binary key sequence which contains `1's and `0's only. Finally, they apply the compressive sensing-based reconciliation method as~\cite{lin2019h2b} to correct the mismatches between their keys.

The authors implemented the proposed system using off-the-shelf piezoelectric energy harvesting products and evaluated its performance with data collected from 24 subjects wearing the devices on different body locations including head, torso and hands. The results showed that KEHKey was able to generate keys at a speed of 12.57 bit/s while reducing energy consumption by $59\%$ compared to accelerometer-based approaches. Additionally, they demonstrated that KEHKey can successfully withstand typical adversarial attacks. Particularly, KEHKey is found to be more resilient to video side channel attacks than its accelerometer-based counterparts.


\subsubsection{Magnetometer} 
Jin et al.~\cite{jin2014magpairing,jin2015magpairing} proposed magnetometer-based system MagPairing to pair two smartphones in close proximity. When two smartphones are in close proximity, their magnetometer readings are highly correlated. The working flow of MagPairing is very similar to ShaVe proposed in Shake well before use~\cite{mayrhofer2009shake} . First, Alice and Bob use D-H protocol to establish an unauthenticated communication channel and Interlock protocol to counter MITM attack. Then, they exchange their magnetic fields readings. If their similarity is higher than a threshold, the DH key will be used to secure the
communication. In the evaluation, the authors implemented MagPairing on several smartphones such as Google Nexus 5 and
GALAXY 3. Evaluation results showed that MagPairing can pair two devices within 4.5 s with more than $90\%$ success rate.

\subsubsection{Electromyogram (EMG)} 
Yang et al.~\cite{yang2016secret} proposed an EMG-based pairing system which is called EMG-KEY. EMG-KEY utilised the electrical activity caused by human muscle contraction (i.e., EMG) to generate secret keys for two wearable devices. MEG-KEY works as follows. First, two wearable devices record raw EMG signal independently when the user performed a random gesture. Then, the EMG signal was rectified and converted into binary bits based on a shape-based key generation method. Finally, the same key was obtained by using Golay Code G(23, 12) in the reconciliation to correct the mismatches. Evaluation results on 10 subjects suggested that EMG-KEY can reach a high bit rate of 5.51 bit/s with a success rate of $88.84\%$.

\subsubsection{Electrode} 
Roeschlin et al.~\cite{roeschlin2018device} proposed a device pairing protocol for two devices that can be physically attached to the body. The suggested protocol is built on the unique body channel: the body channel is used as a part of the pairing protocol to allow two legitimate devices agree on the same secret. Any external devices have no knowledge of the body channel characteristics, and thus cannot guess the secret. In the evaluation, the authors implemented a proof-of-concept and recruited 15 volunteers to verify their idea. Evaluation results showed that the system can achieve 94.3$\%$-99.6$\%$ accuracy.

\subsection{Camera}
As camera are ubiquitous in almost all smartphones, researchers have designed many key generation approaches taking advantage of the cameras in the devices. Below, we introduce several representative works.

McCune et al.~\cite{mccune2005seeing} proposed an authenticated key exchange scheme named Seeing-Is-Believing (SiB). One camera-equipped device can take a photo of a 2D bar code which encodes the cryptographic key of another device. By changing the roles and repeating the above process, both devices can achieve authenticated key exchange. The authors compared their scheme with other key exchange systems using other channels such as ultrasound, audio etc. They claimed that SiB is more secure and convenient to use because the user can identify the device to be paired visually.

However, SiB has several limitations. For example, both devices must be equipped with a camera to achieve mutual authentication. The application scenarios of SiB are limited when one device has limited capabilities such as small size and display. To overcome these drawbacks, Saxena et al.~\cite{saxena2006secure} proposed several extensions to SiB. The proposed approach can achieve mutual authentication when only one device is camera-equipped. This is done by having both Alice and Bob compute a common checksum on public data, then compare their results via a unidirectional visual channel.

Buhan et al.~\cite{buhan2007secure} proposed a biometrics-based secure device pairing system, SAFE. SAFE used cameras to take a picture of another user's face. Then, they applied fuzzy extractor~\cite{dodis2008fuzzy} to extract secret keys from face images. By exchanging some information, both Alice and Bob can agree on the same key. The authors demonstrated that the proposed system is resilient to eavesdropping attack and MITM attack.

The above methods are restricted to two-parties key exchange. GAnGs proposed by Chen et al.~\cite{chen2008gangs} and SPATE proposed by Lin et al.~\cite{lin2010spate} extended SiB to support group device pairing. GAnGs allows a group of devices to collect and distribute authentic information by displays and cameras equipped on each device. SPATE requires users to compare images on different devices to achieve group pairing. Compared to SiB and GAnGs, SPATE achieves better performance in terms of efficiency, accuracy and user experience. The common feature of SiB, GAnGs and SPATE is that they all adopt the barcode format designed by Rohs and Gfeller~\cite{rohs2004using}.

In these methods, cameras are used as OOB visual channel to convey some secret information between intended parties. The application scenarios of these approaches, however, are limited because of the unavailability of cameras in most IoT devices. Additionally, these approaches require human computer interaction to facilitate device pairing process. For example, in the work of McCune et al.~\cite{mccune2005seeing} and Saxena et al.~\cite{saxena2006secure}, the authentication is completed by user answering affirmatively in the last step.

\subsection{Hybrid Approaches}
The above context-based pairing approaches improve the efficiency of pairing IoT devices by removing any human interference in pairing. Using on-board sensors with the same sensing modalities, it is possible to capture a specific physical background such as physical layer characteristics of the same wireless channel, the motion signal of the same user, and ambient sound from the same event. Nonetheless, due to the heterogeneity of IoT devices it is impractical to presume, in some scenarios, that all devices share a common modality of sensing. This limitation sparks some new studies that aim to pair heterogeneous IoT devices by exploiting different sensor modalities. Below, we introduce several representative works belonging to this subcategory.

Miettinen et al.~\cite{miettinen2014context} proposed a context-based zero-interaction pairing protocol for wearable devices by using ambient audio and luminosity. The suggested system utilises context fingerprints to evolve the generated key which is only possible when two devices are in close proximity over a long period of time. Intuitively, it is based on the fact that devices belong to the same user can observe similar context information in a longer time than other devices. Although the proposed scheme does not involve user interaction, it takes a long time to accumulate sufficient entropy to complete authentication. 

Han et al.~\cite{han2018you} proposed a novel context-based pairing scheme named Perceptio for IoT devices that are equipped with different types of sensors. Perceptio utilises time as the common factor across different types of sensors. It produces event fingerprints that can be compared across a range of IoT devices by concentrating on the event timing rather than the individual event sensor data. The principle of Perceptio is that devices co-located inside a physically protected boundary (e.g., smart home) will be able to detect more typical events over time as opposed to outsiders. The authors deployed a variety of heterogeneous IoT devices in a room such as geophone, microphone, accelerometer, motion sensor and a power meter. Evaluation results showed that the legitimate devices in the same room can achieve an average
fingerprint similarity of $94.9\%$ while an outside attacker can only yield $68.9\%$ similarity.

Additionally, some other systems uses different combinations of sensor modalities to pair devices. To name a few, Shrestha et al.~\cite{shrestha2014drone} designed an authentication system by using the combination of four different sensor modalities namely, ambient temperature, precision gas, humidity, and altitude. Liu et al.~\cite{liu2017secure} proposed a pairing scheme for wearable devices based on sound and light. Unisense proposed by Pan et al.~\cite{pan2018universense} and IDIoT proposed by Ruiz et al.~\cite{ruiz2020IDIoT} (the same group of people) leveraged video signals and IMU sensor signals to pair heterogeneous IoT devices.

Key generation schemes using multi-modality sensing present two advantages. On the one hand, it can be used to associate heterogeneous IoT devices that have different sensors, shapes, and sizes. On the other hand, it improves the security of pairing schemes against attackers because it requires the attacker to monitor the various physical environment properties at the same time. Meanwhile, it also has some drawbacks. First, such schemes may take a long time to complete device pairing because the frequency of some events is low (e.g., door closing in~\cite{han2018you}). Second, due to the heterogeneity of sensor data, the processing method is usually more complex and expensive because signals measured in different sensing state-spaces cannot be directly compared. For example, Unisense~\cite{pan2018universense} and IDIoT~\cite{ruiz2020IDIoT} require camera calibration and coordinate transformation to obtain motion signal in the common state-space.
\section{Performance comparison}
\label{sec:performancecomparison}

Because of the diversity of different hardware and protocols, many metrics have been used in prior works. Yet, there lacks a common understanding and sound comparison of existing systems in the current literature. Although several previous works have provided comparisons of some early device pairing schemes~\cite{kobsa2009serial,kumar2009comparative,kainda2009usability}, their studies only provide quantitative comparison by using subjective metrics such as usability~\cite{kumar2009comparative}. In this section, we first summarise commonly used objective metrics in device pairing systems. Then, we provide a comprehensive quantitative comparison of existing schemes. Due to space limitation, we only compare the performance of representative works in each category. Finally, we define five goals of a desirable key generation scheme and provide a qualitative analysis based on this baseline.

\subsection{Objective Metrics and Quantitative Comparison}
\label{subsec:objectivemetrics}

Below, we summarise three commonly used metrics in key generation systems.
\begin{itemize}
    \item Key generation rate (KGR). KGR represents the number of bits in a unit time that a system can generate. It indicates how fast a key generation system can generate or update keys. For the real-time key generation process a high KGR is desirable, as the cryptographic schemes need a certain length of keys. For instance, AES requires a key sequence with a minimum length of 128 bits.
    \item Key agreement rate (KAR). KAR means the number of agreed bits over the total number of generated bits. A high success rate or KAR can considerably improve the efficiency of a device pairing protocol. Ideally, it should be identical for two legitimate devices, and as low as possible for an attacker.
    \item Randomness. Randomness of the key is of upmost importance in a cryptographic protocol because a key with low entropy can be easily cracked by an adversary. In device pairing system, the randomness of the generated keys is evaluated by the popular NIST test suite~\cite{rukhin2001statistical}. The NIST test suite provides 15 tests to evaluate the different randomness features. For example, The purpose of Frequency (Monobit) Test is to evaluate the proportion of zeroes and ones in the key sequence. The goal of Binary Matrix Rank Test is to check for linear dependence among fixed length substrings of the binary key string. The returned result of each test is a \textit{p-value}, which indicates whether the key pass the NIST test or not. Conventionally, if \textit{p-value} $\geq$ 0.01, we say the key passes the corresponding test and it has high randomness.
\end{itemize}

Table~\ref{tab:comparison_performance} summarises the performance of representative works in each category. We have the following observations.
\begin{itemize}
    \item Some studies do not explicitly report the three metrics above, making it hard to achieve a fair comparison with other systems. For example, Shake well before use~\cite{mayrhofer2009shake} used false positive rate and false negative rate in their evaluation. In Miettinen et al.~\cite{miettinen2014context}, the authors evaluated the performance of key revolution in different scenarios but did not give a clear indication about how long it takes to generate a 128 bit key.
    
    \item Some works do not evaluate their systems properly. To be specific, some direct key generation systems do not evaluate the randomness of the extracted keys using NIST test, such as Zeng et al.~\cite{zeng2010exploiting}, Shake-n-Shack~\cite{shen2018shake}, Touch-and-Guard~\cite{wang2016touch}. While in TDS~\cite{xi2016instant}, the authors suggested that their key can be generated by any sophisticated key generation method but do not point out which one. Moreover, since the key is generated from a random number generator, it is not necessary to apply NIST test to validate the randomness of the key.
    
    \item For wireless key generation, CSI-based systems can improve key generation significantly compared to RSSI-based approaches. Unfortunately, some communication technologies cannot provide CSI such as LoRa. Current works on LoRa~\cite{xu2019lora,ruotsalainen2018towards,ruotsalainen2019experimental} still use coarse-grained channel information such as RSSI. Therefore, a future research direction is investigating how to extract fine-grained channel information to improve KGR.
    
    \item For behavioural and physiological traits based key generation schemes, activities involving large displacements usually produce good results such as walking and shaking. The reason is that the SNR of the measured signal is higher as motion signals overwhelm noise when user walks or shakes hands. In contrast, physiological characteristics such as ECG and EMG have limited entropy resulting in low KGR. Moreover, they are not only hard to measure but also contain much noise (i.e., low SNR), which further reduces the performance. But these protocols can be applied in certain scenarios such as implantable IoT devices.
    
    \item Although new sensors provide more possibilities and options to pair various IoT devices, the KGR and KAR is relatively lower than their counterparts such as radio-based and accelerometer-based approaches. Therefore, more research efforts are required to further improve the performance of device pairing schemes using new sensors. However, as these new sensors are not available on off-the-shelf IoT devices and usually require customised hardware or prototype, these obstacles may hinder the progress of research in this direction.
\end{itemize}
\begin{table}[ht]
\small
\caption{A quantitative comparison of different key generation schemes (-- not available).}
\label{tab:comparison_performance}
\scalebox{0.88}{
\begin{tabular}{cccccccc}
\toprule
Source                                                                 & \begin{tabular}[c]{@{}c@{}}Communication\\ Technology/Sensors\end{tabular}                                           & Features         & Literature             & \begin{tabular}[c]{@{}c@{}}Direct\\ Key generation\end{tabular} & \begin{tabular}[c]{@{}c@{}}KGR \\ (bit/s)\end{tabular} & \begin{tabular}[c]{@{}c@{}}KAR\\ (\%)\end{tabular} & Randomness \\ \hline
\multirow{10}{*}{Radio}                                                & ZigBee                                                                                                               & RSSI             & Patwari et al.~\cite{patwari2010high} & $\checkmark$                                                               & 10-22                                                  & 97.8-99.46                                         & $\checkmark$          \\ \cline{2-8} 
                                                                       & Wi-Fi                                                                                                                & RSSI             & Zeng et al.~\cite{zeng2010exploiting}             & $\checkmark$                                                               & 10                                                     & 90                                                 & $\times$          \\ \cline{2-8} 
                                                                       & Wi-Fi                                                                                                                & CIR              & Radio-telepathy~\cite{mathur2008radio}        & $\checkmark$                                                               & 1.17                                                   & >84.15                               & $\checkmark$          \\ \cline{2-8} 
                                                                       & Wi-Fi                                                                                                                & CSI              & TDS~\cite{xi2016instant}                    & $\times$                                                               & 90-120                                                 & 96.5-98                                            & $\checkmark$          \\ \cline{2-8} 
                                                                       & Bluetooth                                                                                                            & RSSI             &    Premnath et al.~\cite{premnath2014secret} & $\checkmark$ & -- & >79                             &   $\checkmark$  \\ \cline{2-8} 
                                                                       & LoRa                                                                                                                 & RSSI             & LoRa-Key~\cite{xu2019lora}               & $\checkmark$                                                               & 18-31                                                  & 98-100                                             & $\checkmark$          \\ \cline{2-8} 
                                                                       & LoRa                                                                                                                 & RSSI             & Ruotsalainen et al.~\cite{ruotsalainen2019experimental}    & $\checkmark$ & -- & 71 - 85 & $\checkmark$            \\ \cline{1-8} 
\multirow{2}{*}{Audio}                                                 & \multirow{2}{*}{Mircophone/Speaker}                                                                                  & Channel response & GeneWave~\cite{xie2017genewave,xie2018genewave}               & $\times$                                                               & 1024                                                   & 88-100                                             & $\times$          \\ \cline{3-8} 
                                                                       &                                                                                                                      & Channel tap      & FREE~\cite{lu2019free}                   & $\checkmark$                                                               & 100-260                                                & 97-100                                             & $\checkmark$          \\ \hline
\multirow{7}{*}{IMU sensors}                                           & \multirow{6}{*}{Accelerometer}                                                                                       & Shake            & Shake well before use~\cite{mayrhofer2009shake}  & \begin{tabular}[c]{@{}c@{}}ShaVe $\times$\\ ShaKe $\checkmark$\end{tabular}       & --                                                     & --                                                 & $\times$          \\ \cline{3-8} 
                                                                       &                                                                                                                      & Shake            & Shake-n-Shack~\cite{shen2018shake}          & $\checkmark$                                                               & $\approx$ 98.4                                                  & 100                                                & $\times$          \\ \cline{3-8} 
                                                                       &                                                                                                                      & Walk             & Gait-Key~\cite{xu2017gait}               & $\checkmark$                                                               & 28                                                     & 97-100                                             & $\checkmark$          \\ \cline{3-8} 
                                                                       &                                                                                                                      & Walk             & Auto-Key~\cite{wu2020auto}               & $\checkmark$                                                               & 3.84-11                                                & 90-100                                             & $\checkmark$          \\ \cline{3-8} 
                                                                       &                                                                                                                      & Walk             & Ravedigar et al.~\cite{revadigar2017accelerometer}                 & $\checkmark$                                                               & 750                                                    & 80-100                                             & $\checkmark$          \\ \cline{3-8} 
                                                                       &                                                                                                                      & Gesture          & Touch-and-Guard~\cite{wang2016touch}        & $\checkmark$                                                               & 7.84                                                   & 98-100                                             & $\times$          \\ \cline{2-8} 
                                                                       & Mag                                                                                                                  & Movement         & Magpairing~\cite{jin2014magpairing,jin2015magpairing}                & $\times$                                                               & --                                                     & >90                                   & $\times$          \\ \hline
\multirow{5}{*}{
Miscellaneous} & \multirow{2}{*}{Piezoelectric sensor}                                                                                & Walk             & KEHKey~\cite{lin2020kehkey}                 & $\checkmark$                                                               & 12.57                                                  & 85-100                                             & $\checkmark$          \\ \cline{3-8} 
                                                                       &                                                                                                                      & Heartbeat        & H2B~\cite{lin2019h2b}                    & $\checkmark$                                                               & 3                                                      & 95.6                                               & $\checkmark$          \\ \cline{2-8} 
                                                                       & ECG                                                                                                                  & Heartbeat        & IMDGuard~\cite{xu2011imdguard}               & $\checkmark$                                                               & 2.84                                                   & 92-100                                             & $\checkmark$          \\ \cline{2-8} 
                                                                       & EMG                                                                                                                  & Hand resonance   & EMG-Key~\cite{yang2016secret}                & $\checkmark$                                                               & 5.51                                                   & 88.84                                              & $\checkmark$          \\ \cline{2-8} 
                                                                       & Electrode                                                                                                            & Body channel     & Roeschlin et al.~\cite{roeschlin2018device}               & $\times$                                                               & --                                                     & 94.3-99.3                                          & $\times$          \\ \hline
\multirow{8}{*}{Hybrid}       & \begin{tabular}[c]{@{}c@{}}Microphone\\ Light sensor\end{tabular}                                                    & --               & Miettinen et al.~\cite{miettinen2014context}       & $\times$                                                               & --                                                     & --                                                 & $\times$          \\ \cline{2-8}
                                                                       & \begin{tabular}[c]{@{}c@{}}Geophone\\ Microphone\\ Accelerometer\\ Infrared motion sensor\\ Power meter\end{tabular} & --               & Perceptio~\cite{han2018you}              & $\times$                                                               & --                                                     & 94.9                                               & $\times$          \\ \cline{2-8}
                                                                       & Camera+IMU                                                                                                           & Gesture          & IDIoT~\cite{ruiz2020IDIoT}                  & $\times$                                                               & --                                                     & 92.2                                               & $\times$          \\ \bottomrule
\end{tabular}
}
\end{table}

\subsection{Subjective Metrics and Qualitative Comparison}
In this paper, we use the following five subjective metrics to evaluate a key generation system for IoT devices. An ideal key agreement system should achieve high performance in each metric, i.e., they should achieve high robustness, usability, practicability, ubiquity and security.
\begin{itemize}
    \item Robustness. It refers to whether a key generation protocol can achieve a high KAR, success rate or true positive rate.
    \item Usability. It refers to the performance of the device pairing system in terms of KGR, completion time, the degree of user involvement and user's ease-of-use perception~\cite{kumar2009comparative}.
    \item Practicability. It refers to the viability of a key generation scheme and is measured by whether the system can work without replying on special hardware.
    \item Ubiquity. It represents the environments a key generation system is applicable to. If a system can work in a variety of environments, it has high ubiquity, and vice versa.
    \item Security. It measures the security of a key agreement protocol against various attacks. The more attacks a key generation system can defend, the more secure it is.
\end{itemize}

Based on the definition above, we provide a qualitative comparison for existing systems and summarise the results in Table~\ref{tab:comparison_performance2}. It should be noted that comparison between different systems is not straightforward and challenging, and our goal is to provide users with a general understanding of the advantages and disadvantages of each technique.  From Table~\ref{tab:comparison_performance2}, we can arrive at a conclusion that no universal approach exists, despite that different studies analyse key generation systems from different perspectives. A system suitable for one scenario may not be applicable to another environment. Meanwhile, there is always a trade-off between different requirements such as security and usability. For instance, a system easy to use is also easy to attack in the meantime. That is to say, a user needs to make efforts to enable a more secure system  by participating the key agreement process (high degree of user involvement). 
\begin{table}[!t]
\centering
\small
\caption{A qualitative comparison of different key generation systems (~\ding{109}--low,~\ding{119}--medium,~\ding{108}--high).}
\label{tab:comparison_performance2}
\begin{tabular}{cccccc}
\toprule
            & Robustness & Usability & Practicability & Ubiquity & Security \\ \hline
Radio       & \ding{108}          & \ding{108}         & \ding{108}              & \ding{119}        & \ding{119}        \\ \hline
Audio       & \ding{108}          & \ding{109}         & \ding{119}              &\ding{119}       & \ding{119}        \\ \hline
IMU sensors & \ding{108}          & \ding{109}         & \ding{108}             & \ding{119}        & \ding{119}        \\ \hline
Camera       & \ding{108}          & \ding{119}         & \ding{109}             & \ding{109}        & \ding{108}        \\ \hline
Miscellaneous sensors & \ding{119}          & \ding{109}         & \ding{109}              & \ding{109}        & \ding{108}        \\ \hline
Hybrid approachs      & \ding{109}          & \ding{109}         & \ding{109}              & \ding{109}        & \ding{108}        \\ \bottomrule
\end{tabular}
\vspace{-0.1in}
\end{table}

Overall, radio-based key generation system is a good choice because it achieves high robustness, usability, practicability and ubiquity. Although the security of such system is medium due to the broadcast nature of wireless radio link, radio-based schemes apply to a variety of IoT devices as most devices possess wireless communication functionalities. 
Audio-based systems suffer from low transmission distance and depend on microphone and speaker which are not always available on IoT devices. The usability and practicability, therefore, are relatively low. The usability of IMU sensor based schemes is low because they also require users to perform some actions such as walking, shaking, bumping to create some common context information. But their practicability is high thanks to the wide availability of IMU sensors on modern IoT devices. Miscellaneous sensors based systems, due to their dependence on special hardware, suffer from low usability, practicability and ubiquity. However, the security in turn benefits from these drawbacks. This is because the signals of such miscellaneous sensors such as user's biometrics are usually hard to copy, fabricate, mimic and eavesdrop. Similar to miscellaneous sensors, multi-modality sensing approaches also achieve low usability, practicability and ubiquity. Additionally, the survey results in Table.~\ref{tab:comparison_performance2} point out some future research directions because it reveals the research gap in each category. For example, the robustness of miscellaneous sensor and multi-modality sensing approaches need more research input.

\section{Attacks and Countermeasures}
\label{sec:attack_countermeaures}
Due to the open nature of wireless communications, attacker's ability of observing users, and side channel information leakage, there are various attacks on an IoT key generation system. In this section, we summarise these attacks and discuss possible countermeasures to address these security issues. Because of the diversity in hardware and design principles, different key generation systems face different attacks. Therefore, we only focus on attacks encountered by most key generation systems. 

We omit some common attacks that have been well investigated in IoT security such as MITM attack, replay attack, DOS attack etc. These attacks have been well examined and many approaches have been proposed to detect and counter such attacks. For example, a MITM attack can be solved by interlock protocol~\cite{rivest1984expose} or message authentication code (MAC). A replay attack can be addressed by using nounces, timestamps or tagging each message with a session ID~\cite{malladi2002preventing}. There are numerous studies in the literature to detect and prevent DOS attack such as~\cite{walters2007wireless,wang2006survey}. For a comprehensive survey of attacks in IoT, interested readers can refer to~\cite{deogirikar2017security,yang2017survey}.

\subsection{Eavesdropping Attack} 
Eavesdropping attack is one of the typical attacking approaches in wireless networks. It has received much attention because many adversarial attacks often follow the eavesdropping activity, such as the MITM attack and the hear-and-fire attack~\cite{kao2006eavesdropping}. 
The attacker, Eve, can simply sit somewhere in the propagation path and eavesdrop all the relevant network traffic for later analysis. Hence, it is easy to perform and hard to detect. 
    
In a key generation protocol, Alice and Bob often need to exchange some information to correct the mismatches between their initial keys (this step is often called information reconciliation). So Eve can eavesdrop this information and utilise it to improve the success rate. There are two ways to address this issue. 
\begin{itemize}
	\item Design a key generation protocol that does not exchange any information. For instance, Ali et al.~\cite{ali2012zero} proposed a zero reconciliation protocol for wearable   devices. To eliminate reconciliation, they first proposed a filtering scheme to significantly improve RSSI correlation between the two communication parties without reducing entropy. Then, they designed an approach that can ensure near-perfect key agreement. The proposed system can achieve $99.8\%$ key matching rate but the KGR is extremely low (only 0.057--0.141 bit/s).
  \item Key generation protocol needs to ensure that Eve still cannot recover the correct key  even with information exchange. Most key generation systems adopt this strategy but in different ways. For example, Xu et al.~\cite{xu2016walkie} exchanged the indices of the samples used for key generation. With the eavesdropped information, the attacker can at most achieve $60\%$ KAR. In~\cite{xu2019lora,lin2019h2b}, the authors used compressive sensing to compress the generated keys into a lower dimension space. Even Eve intercepts the compressed signal, she cannot recover the original key due to the secrecy of compressive sensing theory~\cite{rachlin2008secrecy}.
\end{itemize}
    
Apart from the eavesdropper in radio-based key generation, there are other types of eavesdroppers such as accelerometer eavesdropper in accelerometer-based system~\cite{wang2016touch} and audio eavesdropper in audio-based system~\cite{lu2019free}. Most studies find that if the distance between the eavesdropper and legitimate device exceeds a certain threshold, she cannot use her measurements to generate the same key. Therefore, although eavesdropping attack is easy to perform, most key generation systems can defend it by careful design.

\subsection{Predictable Channel Attack} Predictable channel attack means attackers can perform some regular actions intentionally to cause predictable changes in the received signal of legitimate devices. Jana et al.~\cite{jana2009effectiveness} first identified this attack and later it is found that the majority of RSSI-based key generation systems suffer from such attack. Unfortunately, the majority of RSSI-based key generation systems omit this attack in their evaluation such as~\cite{revadigar2015dlink,xu2019lora,zhang2018channel,zhang2016experimental,ruotsalainen2019experimental,Ali2014Eliminating}. Recently, researchers found that CSI can counter this attack because it can extract channel information from different subcarriers and attacker's movement has different impact on different subcarriers~\cite{liu2013fast,xi2016instant}. To counter predictable channel attack for RSSI-based key generation system, researchers have proposed a variety of methods by introducing different kinds of randomness into the channels, such as using multiple attennas~\cite{ruotsalainen2019experimental}, rotating attenna~\cite{wang2020mobikey}, transmitting random signals~\cite{wang2020practical}, and other schemes~\cite{chen2013smokegrenade,gollakota2011physical,madiseh2012applying,huang2013fast,guillaume2015secret}.

\subsection{Mimicking Attack} 
In the mimicking attack, Eve has the ability to observe how the user is walking, shaking and moving. Then she can repeat user's actions with the aim of generating a similar signal, so that she can extract the same key. Mimicking attack is a major threat to motion-based key generation systems because user's motions are easy to observe and intimate. Although a large portion of these systems have verified the security against mimicking attacks such as~\cite{mayrhofer2009shake,xu2016walkie,xu2017gait,shen2018shake,shen2020securing}. The evaluation is conducted in a controlled environment and using small size dataset. For example,  Shake well before use~\cite{mayrhofer2009shake} recruited 30 participants while Walkie-Talkie~\cite{xu2016walkie} and Shake-n-Shack~\cite{shen2018shake} recruited 20 volunteers only.
    
    Traditionally, it is believed that user's biometrics are hard to copy and fabricate. However, recent studies found that attackers can hack a biometics-based authentication system by generating highly similar biometrics. For example, Eberz et al.~\cite{eberz2017broken} developed an ECG synthesise system to spoof an ECG-based authentication system. The synthetic ECG signals are very close to the ECG of a benign user and the hacking success rate is as high as $81\%$. The feasibility of using synthetic ECG signal to hack an ECG-based key generation system such as~\cite{lin2019h2b,rostami2013heart} has not been investigated yet. But this technology clearly poses a threat to ECG-based key generation systems. 
    
    To mitigate the mimicking attack, we can use signals from multiple sensors such as~\cite{han2018you,liu2017secure}. It is hard for Eve to obtain similar signals from different sensors by eavesdropping and mimicking. But the cost and practicability of using multi-modality sensing is a concern.  
    
\subsection{Side Channel Attack} 
Attackers are becoming more and more powerful with the development of technology. For example,  Davis et al.~\cite{davis2014visual} found that when an audio signal hits an object, it will cause minute vibrations which can be measured by high-speed camera. They proposed a system to recover the sound from a video. With this technology, attackers can hack vibration or acoustic based key generation systems such as~\cite{wang2016touch,lu2019free,karapanos2015sound}. Recent studies also showed that videos can be used to recover the motion data of people's activities~\cite{ye2017cracking}. Therefore, motion or vibration-based key generation schemes are vulnerable to a video analysis attack. In fact, Bruesch et al. ~\cite{bruesch2019security} have demonstrated the feasibility of using video to estimate user's gait signals. To defend against such attacks, one solution is to introduce random signal in the key generation process as how researchers did to counter eavesdropping attack above. Another possible solution is to borrow the idea from face recognition with liveness detection system~\cite{bao2009liveness}. The purpose of liveness detection is to detect whether a face is “alive” or just a fraudulent reproduction. Therefore, by incorporating liveness detection, a key generation system can defend such attacks when they reproduce a similar signal from side channel information.

\subsection{Summary}
The vulnerability of key generation systems subject to different attacks is summarised in Table~\ref{tab:attacks}. We can see that different key generation systems have varied vulnerability levels to attacks.
\begin{table}[!t]
\centering
\small
\caption{Vulnerability of key generation systems to different attacks. (~\ding{109}--low,~\ding{119}--medium,~\ding{108}--high).}
\label{tab:attacks}
\begin{tabular}{lcccc}
\toprule
\multicolumn{1}{c}{}     & \begin{tabular}[c]{@{}c@{}}Eavesdropping\\ Attack\end{tabular} & \begin{tabular}[c]{@{}c@{}}Predictable Channel\\ Attack\end{tabular} & \begin{tabular}[c]{@{}c@{}}Mimicking\\ Attack\end{tabular} & \begin{tabular}[c]{@{}c@{}}Side channel\\ Attack\end{tabular} \\ \hline
Radio       & \ding{108}          & \ding{108}         & \ding{119}              & \ding{119}               \\ \hline
Audio       & \ding{108}          & \ding{109}         & \ding{119}              &\ding{109}              \\ \hline
IMU sensors & \ding{108}          & \ding{119}         & \ding{108}             & \ding{119}               \\ \hline
Camera       & \ding{108}          & \ding{119}         & \ding{108}             & \ding{119}           \\ \hline
Miscellaneous sensors & \ding{119}          & \ding{109}         & \ding{119}              & \ding{109}              \\ \hline
Hybrid approaches      & \ding{119}          & \ding{119}         & \ding{109}              & \ding{109}    \\ \bottomrule
\end{tabular}
\vspace{-0.1in}
\end{table}

The rapid development of technology is a double-edged sword. On the one hand, it brings more opportunities for researchers to design novel IoT key generation systems by using novel algorithms (e.g., deep learning~\cite{wu2020auto}) and hardware (e.g., bio-sensors~\cite{roeschlin2018device}). On the other hand, attackers can use more powerful devices (e.g., high-speed camera~\cite{davis2014visual}) to perform more advanced attacks. Therefore, in addition to the above well studied attacks, researchers should consider potential attacks especially that exploit side channel information leakage. Additionally, when we survey the literature, we find that almost all the studies focus on developing novel key generation systems but very limited work has been done to analyse the vulnerabilities of existing systems. Some challenging studies are hard to conduct but of great value. To name a few, Edman et al.~\cite{edman2011passive} challenged the common assumption in physical layer key generation that if Eve is more than half wavelength away from Alice or Bob, her channel measurements have low correlation with that of Alice and Bob. Instead, they found that there is a strong correlation in Eve's channel measurements which is contradictory to previous results. Bruesch et al. ~\cite{bruesch2019security} analysed the security of several gait-based key generation system and pointed out that there is a bias in the generated keys. Therefore, studies from the perspective of attackers need more research efforts.

In addition to the attacks above, the generated keys may also suffer from poor randomness and key collision. The poor randomness will make the key easy to be hacked, and key collision may lead to duplicated keys in the key generation process. As mentioned in Section~\ref{subsec:objectivemetrics}, key randomness is evaluated by NIST test. However, Table~\ref{tab:comparison_performance} reveals that a large portion of work do not verify the randomness of the generated keys. Therefore, the usability of these schemes in real applications is still questionable. Key collision has been rarely mentioned in the previous studies because the focus of prior work is whether Eve can generate the same key as Alice and Bob rather than if Alice and Bob will generate the same key multiple times. Therefore, key randomness and key collision require more attention in the future research.
\section{Challenges and Directions}
\label{sec:challenges_directions}
So far, we have surveyed recent advances on IoT key generation, compared their performance and analysed various security issues. A number of research challenges still require further investigation. We discuss some of them and point out several future directions in this section.

\begin{itemize}
    \item \textbf{Key Generation for Emerging Technologies.} Previous radio-based key generation works have been mainly focused on short range communications such as Wi-Fi and ZigBee. However, limited efforts have been made in key generation using new wireless communication technologies such as Low power wide area networks (LPWAN) and 5G. LPWAN  have started to prevail in the last few years, such as LoRa/LoRaWAN, NB-IoT. The key challenge for LPWAN is channel reciprocity may not hold any more for those techniques~\cite{xu2019lora}. LoRaWAN protocol specifies a one-second delay between the uplink and downlink transmissions, which will significantly impact the channel reciprocity. NB-IoT employs the FDD duplex mode where the channel reciprocity may not hold at all~\cite{li2018constructing}. Therefore, how to key generate keys in low channel reciprocity scenarios remains an open research problem. Another hot communication technology is 5G which uses several new features such as mmWave, massive MIMO and highly directional beamforming~\cite{yang2015safeguarding}. The current study on 5G key generation still relies on simulation and practical key generation systems have not been developed yet. These new communication technologies bring not only challenges but also opportunities. For instance, high directional beamforming can be used to thwart co-located attackers
in key generation~\cite{jiao2018secret}. We believe that designing key generation systems for these emerging communication technologies will be a hot research direction in the future.

    \item \textbf{Group Key Generation.} With the growing number of IoT devices, it is more common to pair a group of IoT devices in the same network or context. Unfortunately, most existing key generation systems work in peer-to-peer communication mode. While many studies have been conducted in group key generation~\cite{wang2011fast,liu2012collaborative,liu2014group,gu2016scalable,xu2016group,revadigar2017accelerometer}, they suffer from either efficiency or practicability issues.  Traditional pairwise key generation protocols cannot be directly extended to group key generation scenarios. This is because the secret information between a pair of legitimate devices cannot be efficiently and securely distributed to other IoT devices. The naive solution is first to apply pairwise key generation schemes to generate a secure channel between each pair of nodes, then exchange some information to agree on the same group key. However, the computation and communication overhead of this approach increases linearly with the group size. Some solutions are impracticable because off-the-shelf IoT devices cannot provide the channel characteristics they used (e.g., phase~\cite{wang2011fast}). Therefore, how to design an efficient and practical group key generation protocol remains an open question.

    \item \textbf{User Friendliness.} From a user-friendliness point of view, the user needs to have minimal involvement during key generation process. Unfortunately, the majority of systems described in this paper require user in the loop. The user plays a crucial role in some key generation systems, such as introducing randomness by shaking~\cite{mayrhofer2009shake,shen2018shake} or walking~\cite{xu2016walkie,xu2017gait}, comparing results~\cite{mccune2005seeing,saxena2006secure}. Even for radio-based key generation when two devices are static, the channel variance is caused by moving subjects or objects in the environment. This is why several previous surveys classify key generation systems based on HCI~\cite{chong2014survey,yang2017survey}. For example, Chong et al.~\cite{chong2014survey} provided an in-depth analysis of device pairing schemes from the perspective of user involvement. A desirable key generation system requires as little user involvement as possible. However, a highly user-friendly key generation system does not come for free but it decreases the performance. For example, the zero-interaction device pairing scheme proposed by Miettinen et al.~\cite{miettinen2014context} does not need user interaction but requires a long time to complete authentication. Therefore, how to design a fast and practical key generation system with little user involvement requires further investigation.
    
    \item \textbf{Balancing Different Design Requirements.} While the battery life and processing capability of IoT devices have been greatly improved over the past few years, some categories of devices are still limited by on-board resources such as wearable and IMD devices. Resource constraints have been long identified as the primary challenge in designing an IoT system and researchers have made significant efforts and progress to find a balance between performance and resource consumption. With regards to key generation protocols, this trade-off becomes more complex. As mentioned earlier, there are several requirements of a desirable IoT key generation protocol such as security, efficiency and usability. There is a fundamental tension between these goals, particularly when examining these goals in the context of realistic application scenarios. Unfortunately,  finding a suitable balance between these tensions is non-trivial. Therefore, we believe that the conflicting requirements of security, reliability, and usability in IoT devices will still be a research problem in the near future and motivate more inventive works.
  
    \item \textbf{Theory vs Practice.} Many studies are based on theoretical analysis and simulation evaluation~\cite{wang2011fast,zhang2016efficient,zhang2019design,xu2016group}, while many other studies are based on real-world measurements using off-the-shelf hardware~\cite{ruotsalainen2019experimental,xu2016walkie,xu2017gait,jin2015magpairing,shen2018shake,findling2014shakeunlock}. Only a small portion of studies in the literature present both theoretical analysis and practical validation such as~\cite{xu2019lora,cai2011good}. This is probably because the results obtained from theory may not apply to real-world measurements directly. For example, according to wireless communication theory~\cite{goldsmith2005wireless}, the channel will be statistically uncorrelated if two devices are separated by half wavelength away. In practice, however, this is not always true because of the poor multi-path conditions and interference in some environments~\cite{revadigar2015dlink,edman2011passive}. As another example, although channel phase randomness has been extensively investigated for key generation~\cite{wilson2007channel,wang2011fast}, it is only validated in simulation. Unfortunately, off-the-shelf devices cannot provide accurate channel phase information due to noise and offset~\cite{zhang2019calibrating}. Therefore, a solid and rigorous work needs both theory support and field validation.  Unfortunately, this gap has not been filled so far.
    
\end{itemize}

\section{Conclusion}
\label{sec:conclusion}
IoT is seen as the future of the world. Over the past decade, a large variety of IoT devices have hit the market. Therefore, ensuring secure communication in IoT system is of the upmost importance. Accordingly, the number of research articles related to secure D2D communication has been increasing exponentially. A large number of key generation systems have been proposed and designed. In this paper, we have reviewed, analysed, and compared recent solutions. Based on a novel taxonomy, we categorised existing works based on the hardware interface used in the device pairing systems. Moreover, we analysed the security of current key generation systems and pointed out several potential future research directions. Although we have reviewed more than 100 articles in this survey, the list of existing systems is by no means exhaustive but covers the majority of recent advances and directions. We hope this survey can help researchers identify research gaps and find solutions easily. We also would like to invite all researchers to broaden this exciting area and provide new insights.

\begin{acks}
This work is supported by the APRC grant (Project No. 9610485) and the Start-up grant (Project No. 7200642) from City University of Hong Kong.
\end{acks}

\bibliographystyle{ACM-Reference-Format}
\bibliography{sample-base}

\end{document}